\documentclass[preprint,12pt]{elsarticle}
\usepackage[printonlyused]{acronym}
\usepackage{hyperref}
\usepackage[font=small,labelfont=bf, justification=centering]{caption}
\usepackage{mathtools}
\usepackage{bm}
\usepackage[section]{placeins}
\usepackage{fixltx2e}
\DeclarePairedDelimiter\ceil{\lceil}{\rceil}
\usepackage{subfigure}
\usepackage{xcolor}

\usepackage{amssymb}

\makeatletter
\def\ps@pprintTitle{%
  \let\@oddhead\@empty
  \let\@evenhead\@empty
  \let\@oddfoot\@empty
  \let\@evenfoot\@oddfoot
}
\makeatother

\begin{document}

\begin{frontmatter}


\title{Localization in Power-Constrained Terahertz-Operating Software-Defined Metamaterials}


\author[label1,label2]{\vspace{-3mm}Filip Lemic\corref{cor1}}
\cortext[cor1]{Corresponding author: \url{filip.lemic@uantwerpen.be} (Filip Lemic)}
\author[label2]{Sergi Abadal}
\author[label3]{Chong Han}
\author[label1]{Johann M. Marquez-Barja}
\author[label2]{Eduard Alarc\'{o}n}
\author[label1]{Jeroen Famaey}
\address[label1]{Internet Technology and Data Science Lab, University of Antwerp - imec, Belgium}
\address[label2]{NaNoNetworking Center in Catalunya, Polytechnic University of Catalunya, Spain}
\address[label3]{Terahertz Wireless Communications Lab, Shanghai Jiao Tong University, China\vspace{-5mm}}

\begin{abstract}
\noindent
Software-Defined Metamaterials (SDMs) show a strong potential for advancing the engineered control of electromagnetic waves. 
As such, they are envisioned to enable a variety of exciting applications, among others in the domains of smart textiles, high-resolution structural monitoring, and sensing in challenging environments. 
Many of the applications envisage deformations of the SDM structures, such as their bending, stretching or rolling, which implies that the locations of metamaterial elements will be changing relative to one another. 
In this paper, we argue that if the metamaterial elements would be accurately localizable, this location information could potentially be utilized for enabling novel SDM applications, as well as for optimizing the control of the elements themselves.  
To enable their localization, we assume that these elements are controlled wirelessly through a Terahertz (THz)-operating nanonetwork. 
We consider the elements to be power-constrained, with their sole powering option being to harvest energy from different environmental sources. 
By means of simulation, we demonstrate sub-millimeter accuracy of the two-way Time of Flight (ToF)-based localization, as well as high availability of the service (i.e., consistently more than 80\% of the time), which is a result of the low energy consumed in the localization process. 
Finally, we qualitatively characterize the latency of the proposed localization service, as well as outline several challenges and future research directions.  

\end{abstract}


\end{frontmatter}


\acrodef{SDMs}{Software-Defined Metamaterials}
\acrodef{SDM}{Software-Defined Metamaterial}
\acrodef{THz}{Terahertz}
\acrodef{UWB}{Ultra Wide-Band}
\acrodef{FPGA}{Field Programmable Gate Array}
\acrodef{ToF}{Time of Flight}
\acrodef{AoA}{Angle of Arrival}
\acrodef{RSS}{Received Signal Strength}
\acrodef{AP}{Anchor Point}
\acrodef{3D}{3-Dimensional}
\acrodef{TS-OOK}{Time-Spread ON-OFF Keying}
\acrodef{RF}{Radio Frequency}
\acrodef{2D}{2-Dimensional}
\acrodef{GPS}{Global Positioning System}
\acrodef{IoT}{Internet of Things}
\acrodef{WSN}{Wireless Sensor Network}
\acrodef{LoS}{Line of Sight}

\section{Introduction}
\label{introduction}

Metamaterials are manufactured structures with engineered properties not found in natural materials. 
As such, metamaterials (and metasurfaces, their \ac{2D} counterparts) can support controlled manipulation of electromagnetic waves (e.g., reflection, absorption, scattering)~\cite{lemic2019survey,abadal2020programmable}.
This unprecedented level of control is achieved by carefully designing a periodic array of sub-wavelength metamaterial elements, also known as unit cells~\cite{abadal2020programmable}. 
Until recently, these unit cells have been designed for supporting a single application and/or operational condition, and, therefore, have not been reusable across applications nor operations.
To enable a substantially higher level of controllability of metamaterial elements in real-time, Liaskos \emph{et al.}~\cite{liaskos2018new} proposed \acp{SDM}. 
The SDM paradigm envisions embedding a communication nanonetwork of controllers within the metamaterial or metasurface, with the roles of controllers being i) actuation of the metamaterial elements and/or ii) collection of sensed readings from the elements through a two-way exchange of the sensing/actuation commands and responses between the metamaterial elements and the outside world~\cite{lemic2020idling}.

There are several technological candidates for enabling a nanocommunication network to support the above-mentioned controllability features, including both wired (primarily \acp{FPGA}~\cite{abadal2020programmable}) and wireless solutions (mainly graphene-based \ac{THz} nanonetworks~\cite{lemic2019assessing}).  
Intuitively, both candidates come with strengths and weaknesses, with the primary concern being the wiring and form-factor vs. communication reliability trade-off between wired and wireless solutions~\cite{liaskos2015design,abadal2020programmable,lemic2019assessing}.
The question of which candidate to utilize for enabling the foreseen nanonetwork is currently not fully answered and a subject of ongoing research~\cite{tasolamprou2018intercell,liu2018programmable}. 

SDMs are envisioned to enable a variety of novel applications in a number of domains~\cite{lemic2019survey}.
They will have to be flexible to fully achieve the promise of some of the envisioned applications (more details in Section~\ref{related_works}).
Hence, one of the aims of metamaterial-focused research is to develop SDMs that have the possibility of bending, stretching, and rolling.~\cite{walia2015flexible}.    
In such setups, the locations of the metamaterial elements will be dynamically changing relative to one another.
Intuitively, for such SDM-based applications the location of a given metamaterial element will play a role in the control of that element.   
However, to the best of our knowledge, the question if it is at all possible, as well as how to localize the metamaterial elements relative to one another did not receive any attention in the research community to date.

\subsection{Contributions}

To address the above issues, we start by considering different approaches for traditional localization (i.e., the ones originating from e.g., \ac{WSN} and the \ac{IoT} contexts) utilizing \ac{RF} signals. 
Based on these insights, we argue that the two-way \ac{ToF}-based trilateration is the most suitable candidate for enabling localization in THz-operating SDMs.
Consequently, we demonstrate the feasibility of localizing SDM elements by utilizing RF signals in THz frequencies (i.e., 300~GHz to 10~THz). 
We do that for power-constrained SDMs, where harvesting energy from nearby environmental sources is the sole powering option for the SDM elements.
This is done as it is expected that such power-constrained SDMs will be utilized in the future, primarily to reduce their form-factors by removing the need for excessive wiring~\cite{lemic2019survey}. 
In particular, we demonstrate that the two-way \ac{ToF}-based trilateration for \ac{TS-OOK}, a relatively novel modulation and coding scheme specifically designed for nanocommunication, can achieve a sub-millimeter-level average errors in localization, even for highly constrained SDMs powered solely through harvesting energy from surrounding air-vibrations by exploiting piezoelectric effect of ZnO nanowire-based harvesters.
In addition, we demonstrate consistently high availability of the localization service, resulting from the low energy consumed in the process, showing the applicability of the proposed localization scheme for highly power-constrained \acp{SDM}.  
Further, we characterize the influence of a large number of relevant system parameters along the values expected to be utilized in real-life deployments (e.g., operational bandwidth, energy harvesting rate, SDM spacing, number of localization anchors, etc.) on the localization accuracy and availability. 
Finally, we provide a rule-of-thumb characterization of the latency of performing such localization as a function of different system parameters, and outline several open challenges and future research directions. 

\subsection{Structure}
This paper is structured as follows. In Section~\ref{related_works}, we discuss the related works and efforts pertaining to traditional RF-based localization, followed by outlining several potential benefits of the localization feature in SDMs.
In Section~\ref{system}, we discuss the envisioned system for two-way ToF-based localization in THz-operating and power-constrained SDMs.
In Section~\ref{methodology}, we outline the evaluation methodology, while in Sections~\ref{results} and~\ref{open_challenges} we deliver the results of our evaluation and discusses several open challenges and potential future research directions, respectively.
The paper is concluded in Section~\ref{conclusion}. 

\section{Related Works}
\label{related_works}

\subsection{Localization in Software-Defined Metamaterials: Opportunities}

SDMs are envisioned to serve as an enabler of a variety of applications in different domains~\cite{lemic2019survey,abadal2020programmable}. 
Among others, SDMs are expected to be embedded in smart textiles for supporting features such as wireless touch~\cite{tian2019wireless}.
SDMs are also expected to be used in vehicular communication for enhancing ranging capabilities~\cite{tak2017metamaterial} or noise cancellation~\cite{sato2007metamaterials}. 
Moreover, some SDM applications envision metamaterial-based sensing on rough mobile surfaces (e.g., temperature~\cite{kairm2014concept} and strain sensing~\cite{melik2009flexible}), or sensing in fluids~\cite{labidi2011meta}).
To enable such applications, one of the aims of metamaterial-focused research is to develop flexible SDMs with the possibility of bending, stretching, rolling, etc.~\cite{walia2015flexible}, providing a strong indication that SDMs will eventually be utilized in scenarios with mobility, hence the locations of the metamaterial elements will be dynamically alterable relative to one another.

As mentioned, localization in SDMs is a topic that did not receive any attention from the community to date. 
That is with an exception of our previous work~\cite{lemictoward} (n.b., a conference paper), where we outlined the concept of two-way \ac{ToF}-based trilateration for enabling localization in THz-operating power-constrained SDMs. 
In this work, we extend our initial findings by, among others, evaluating the proposed approach with a substantially higher level of realism (e.g., by considering an SDM-specific and frequency-dependent channel model) and along a more exhaustive set of system parameters (e.g., by considering the effects of the number and constellation of localization anchors and mobility patterns of the nanonodes), as well as by utilizing a more comprehensive set of performance metrics (i.e., localization accuracy and availability, and latency of reporting location estimates).

Given the novelty of the topic at hand, to the best of our knowledge there are no obvious related works along which we could position and benchmark our work against. Therefore, in the following we take a ``what if'' perspective and aim at providing a high-level outline of some potential benefits of the localization feature in SDMs. We categorize these opportunities based on the application and network optimization standpoints. Note that our intention is to present a few motivating examples, in contrast to providing an exhaustive overview of all the potential opportunities.

\subsubsection{Application Perspective}

Taking the application perspective, the possibility of estimating the locations of mobile wireless communication-enabled devices is an enabler of a variety of novel services, traditionally known as location-based services~\cite{mohapatra2005survey,huang2009survey}.
In the context of SDMs, arguably the most intuitive example of the benefits of localization of metamaterial elements comes from the aforementioned fact that some SDMs are envisioned to be used for sensing applications in challenging environments, usually with mobility~\cite{salim2018review,kairm2014concept,melik2009flexible,labidi2011meta}.
It is well-known from traditional \acp{WSN} that the location where a reading is obtained is an important piece of metadata~\cite{karl2007protocols} about the reading itself.
Hence, it is reasonable to assume that in the domain of SDM-based sensing localizable readings would be of a certain value (e.g., wireless touch should be a localizable stimulus). 
SDMs also find their applications in intelligent wireless environments~\cite{tasolamprou2019exploration,di2019smart}. 
Although initially SDMs are in this context expected to be static entities (often attached to walls and ceiling of a deployment environment)~\cite{zhao2020metasurface}, with the progress in their development they could become entities integrated in e.g., person's clothes, thus the person could become an active participant in the intelligent wireless environments.
Introducing a person as an active element could for example benefit the mmWave communication or macro-scale localization capabilities.
This is because a mobile obstacle that the human body represents usually has a negative effect on the performance of these and similar systems~\cite{sur201560,geng2013modeling}. 
Given that eventually humans could become the active elements by means of e.g., smart textiles, it becomes clear that the locations of the SDM elements embedded in such textiles could be used for a more optimized control of the impinging electromagnetic signals based on the locations of the elements themselves, which in turn could be used for fine-grained optimizations of the macro-scale communication and localization capabilities in the intelligent wireless environments. 

\subsubsection{Network Optimization Perspective}

From the network optimization standpoint, it is first worth emphasizing that location-based optimizations represent a promising research domain for optimizing the performance of more traditional wireless networks across their entire protocol stacks~\cite{di2014location}.
In the context of SDMs, the aim of location-based optimizations would be to improve the performance of the wireless nanonetworks for controlling the metamaterial elements (more details in Section~\ref{sec:sdm_context}).  
Generally speaking, the reactivity of any SDM to electromagnetic waves of a given frequency changes as the spacings between its metamaterial elements change~\cite{taghvaee2020radiation,abadal2020programmable}.
By quantifying the level of these changes (i.e., the locations of the metamaterial elements), one could potentially design more advanced approaches for controlling the radiation pattern characteristics of the SDMs.
Similarly, one could develop a mechanism for excluding certain elements from the operation of an SDM for a fraction of time, as they have been e.g., overly stretched (i.e., roughly speaking extended beyond the wavelength of the controlled waves)~\cite{abadal2020programmable}.
Finally, given that SDMs are envisioned to consist of a relatively large number of elements (from hundreds to multiple thousands)~\cite{lemic2019survey}, it is an open question on how to individually address each element in a constrained time-frame. 
In this regard, directing wireless signals from the gateways/controllers toward the metamaterial elements is a promising option~\cite{saeed2019workload}, as it would reduce the interference with other elements compared to omnidirectional communication, thus also reducing the time needed for the control of the elements.
However, under mobility of the SDM, fast directional communication (i.e., without a time-consuming beam-search~\cite{sur201560})  between the controllers and the elements could be feasible if the locations of the elements would be known with a high level of accuracy.
In other words, if the locations are known, the beams could be steered in the \ac{LoS} direction of the elements.   
Such a location-based optimization has been shown to mitigate interference~\cite{halbauer2012interference} and significantly reduce the beam-search latency~\cite{abdelreheem2016millimeter} in macro-scale mmWave systems, which could serve as an additional argument about its utility in the SDM context.

\subsection{Traditional RF-based Localization}

Traditional \ac{RF} localization approaches (i.e., algorithms) can be roughly categorized into fingerprinting, proximity, and geometric-based~\cite{brena2017evolution,oguntala2018indoor}.
Fingerprinting approaches rely on correlating various signal features from an unknown location with a set of respective signal features at pre-surveyed locations~\cite{caso2019vifi}. 
Due to the fact that it is practically impossible to pre-survey signal features at nanoscale, we believe fingerprinting is infeasible for localization in SDMs. 
Proximity-based approaches rely on the proximity between a device to be localized and a set of anchors with known locations. 
These approaches are designed to provide coarse-grained localization~\cite{lemic2015experimental} and are, therefore, not suitable for the desired nanoscale localization with high accuracy.
Geometrical approaches utilize different signal features (mostly \ac{RSS}, \ac{AoA}, and ToF) for estimating distances (or angles) between devices, which then serve as a basis for estimating the unknown locations~\cite{lemic2016localization}.

\begin{table}[t]
\vspace{-5mm}
\centering
\caption{Popular physical layers used in localization~\cite{xiong2015tonetrack}.}
\label{tab:table1}
\small
\begin{tabular}{p{5cm} p{3.0cm} p{3.2cm}}
\textbf{Physical layer}  & \hfil \textbf{Bandwidth}      & \hfil \textbf{Raw resolution} \\ \hline
IEEE~802.11a/g           & \hfil 20~MHz                  & \hfil 15~m                    \\
IEEE~802.11n             & \hfil 40~MHz                  & \hfil 7.5~m                   \\
IEEE~802.11ac            & \hfil $<$160~MHz              & \hfil $>$1.9~m                \\
Ultra WideBand (UWB)     & \hfil $>$500~MHz              & \hfil $<$0.6~m                \\ 
IEEE~802.11ad            & \hfil $>$2~GHz                & \hfil $<$15~cm   			 \\ 
\textbf{Terahertz - macroscale} & \hfil $\bm{>}$\textbf{10~GHz} & \hfil $\bm{<}$\textbf{3~cm}   \\ 
\textbf{Terahertz - nanoscale} & \hfil $\bm{>}$\textbf{1~THz} & \hfil $\bm{<}$\textbf{0.3~mm}   \\ \hline
\end{tabular}
\end{table}

AoAs cannot be estimated without antenna arrays or complex signal processing, both being infeasible for the considered nanonodes with highly constrained capabilities~\cite{rong2006angle}.
RSS is a highly fluctuating signal feature with logarithmic dependence to the distance between devices, hence it is known to be highly inaccurate~\cite{lemic2016localization,lemic2019regression}. 
Estimation of the ToF requires tight synchronization between the communicating devices, which is hardly achievable for the considered resource-constrained nanonodes.  

Luckily, a two-way ToF-based approach removes the synchronization requirement and we, therefore, reason that two-way ToF-based localization can potentially enable localization in SDMs.
For estimating two-way ToF, a signal is transmitted by one device and the time measuring is started. 
After a predefined period upon the reception of the original signal, the other device retransmits the signal. 
Upon the following reception, the first device can estimate the two-way ToF by subtracting the predefined time that passed between the reception of the original signal and the transmission of the new one from the total time passed since the original transmission.   
The accuracy of this method will depend on the sampling constraints resulting from the Nyquist theorem. 
Extending the classification proposed by Xiong~\emph{et al.}~\cite{xiong2015tonetrack}, in Table~\ref{tab:table1} we summarize the raw resolution of different physical layers prominently used for localization, with the raw resolution defined as the speed of light divided by the available bandwidth.
In the THz frequencies, different spectrum windows of at least 10~GHz are unlicensed and ubiquitously available~\cite{boronin2014capacity} for macroscale communication. Moreover, at the nanoscale (i.e., with the communication ranges of less than 50~cm) the available bandwidth is in the range of several THz~\cite{abadal2019graphene}.
The discrepancy between the available bandwidths at macro and nanoscale is primarily due to increase in the molecular absorption with communication distance, as shown in~\cite{abadal2019graphene}.
This yields the theoretically achievable raw resolutions of less than 3~cm and less than 0.3~mm at the macro and nanoscale, respectively, which demonstrates a potential for high accuracy in THz-based localization. 

\section{Localization in SDMs}
\label{system}

\subsection{Software-Defined Metamaterials: Context}
\label{sec:sdm_context}

\begin{figure}[!t]
\centering
\includegraphics[width=0.82\linewidth]{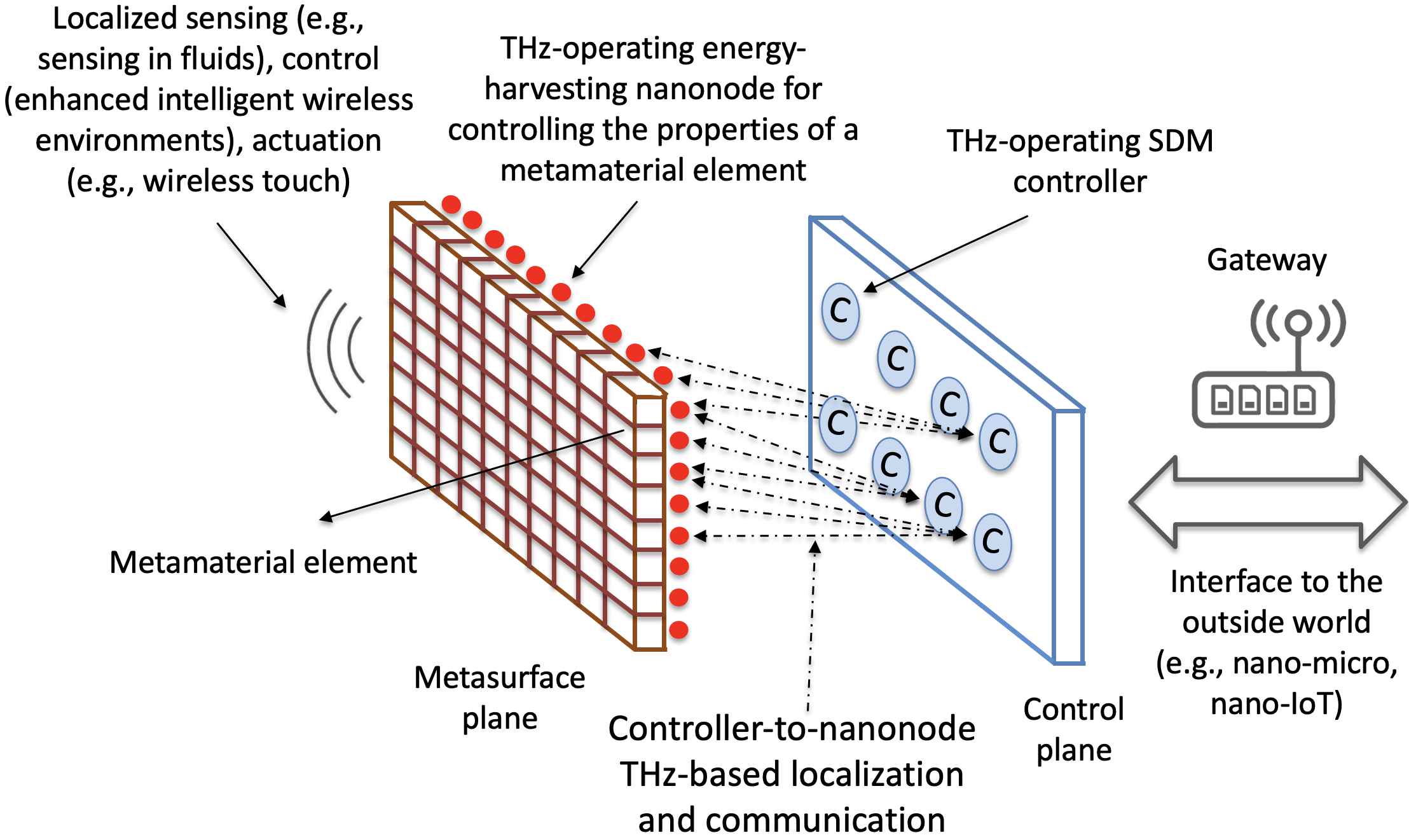}	
\caption{Envisioned SDM architecture for controlling and localizing metamaterial elements}
\label{fig:architecture}
\end{figure}

Figure~\ref{fig:architecture} depicts the envisioned SDM architecture, where each controller can interact with its associated metamaterial elements.
The interaction is supported through power-constrained nanonodes, whose main purpose is two-way wireless communication with the controllers for the functionalities of adjusting the properties of the metamaterial elements and delivering their readings. 
We consider THz-based wireless nanocommunication between the controllers and the power-constrained nanonodes, which allows (compared to a wired solution) for a reduced form factor, however at the cost of highly constrained amount of available energy (i.e., with harvesting energy from environmental sources being the only powering option)~\cite{jornet2013graphene}.
THz-based nanoscale wireless communication between the controllers and the nanonodes can be supported by emerging materials, primarily graphene, and its feasibility has been demonstrated by various works in the literature, e.g.,~\cite{jornet2013graphene,abadal2019graphene}.
Moreover, transceivers for THz-based communication at more than 10~GHz bandwidth, featuring low energy consumption approaching 1 pJ per bit, have been demonstrated within small areas expected in the SDM context (e.g.,~\cite{yu2014architecture}), paving the way toward the envisioned SDM localization.

We use \ac{TS-OOK} as the modulation and coding scheme as it is a \textit{de-facto} standard for THz nanocommunication~\cite{jornet2011information}.
\ac{TS-OOK} is based on the exchange of short pulses (usually 100~fs long~\cite{jornet2011information}) spread in time, where the pulse carries one bit of information, i.e., a logical "1", while a logical ``0'' is represented by silence.  
The time between the transmission/reception of two consecutive bits of a packet is characterized by the ratio $\beta$ between the duration of a TS-OOK pulse and the time between the transmissions/receptions of two consecutive bits of information. 
$\beta$ is usually fixed and much longer (typically two to three orders of magnitude) than the duration of a TS-OOK pulse.
The benefits of THz-based TS-OOK communication include simplicity of transceiver architecture~\cite{jornet2014femtosecond}, extremely low energy consumption in the range of 1~pJ per transmitted bit of information~\cite{jornet2011information} for omni-directional communication ranges of up to 1~m~\cite{abadal2019graphene}, making it a suitable candidate for the realization of the envisioned SDM localization.
This is further accentuated by the fact that THz channel path loss within an SDM should, perhaps counterintuitively, not necessarily be very high because distances are small and the environment is enclosed~\cite{tasolamprou2019exploration}.

\begin{figure}[!t]
\centering
\includegraphics[width=0.75\linewidth]{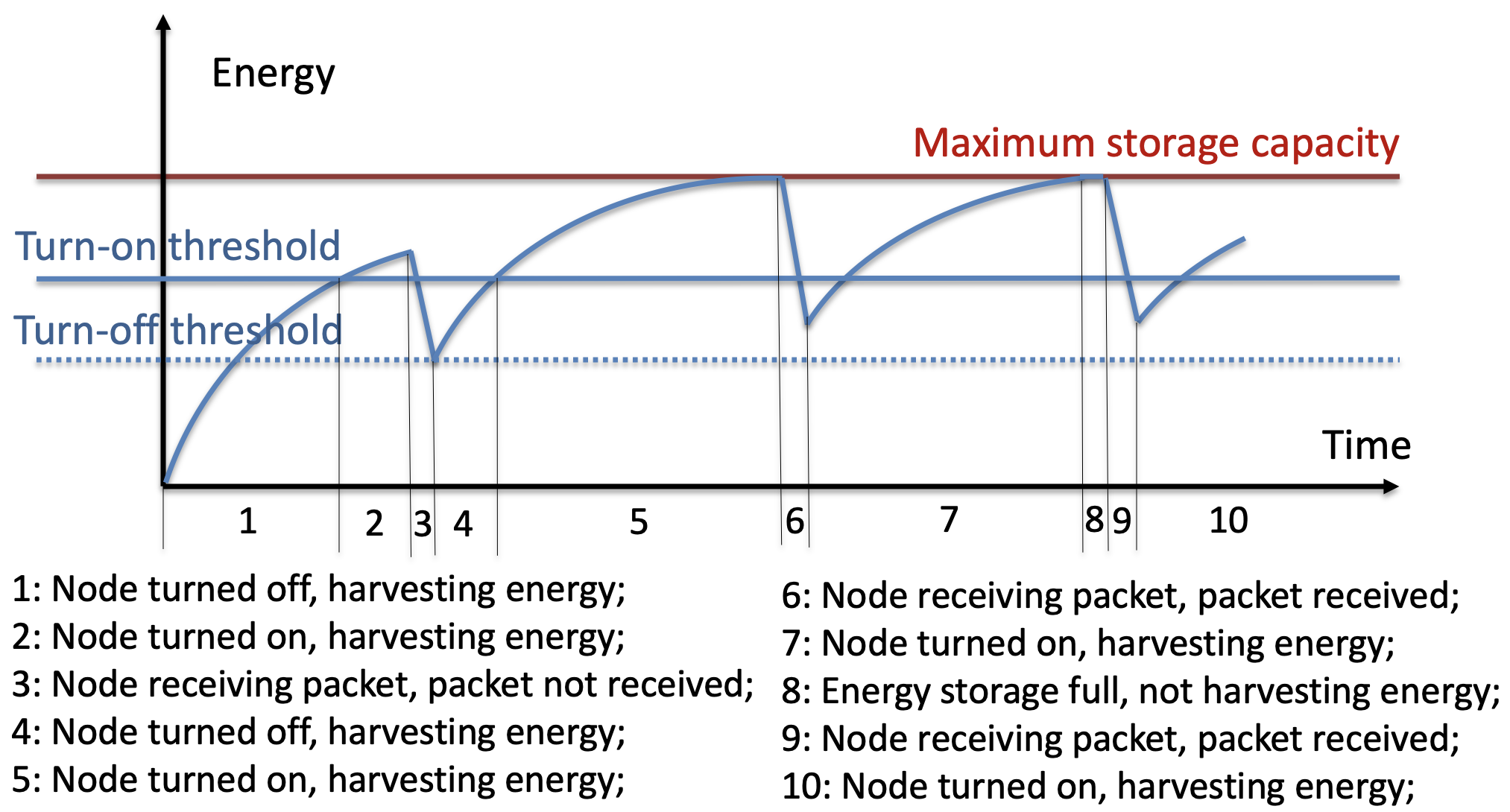}
\caption{Lifecycle of an power-constrained nanonode~\cite{lemic2019assessing}}
\label{fig:intermittent}
\end{figure}

\subsection{Power-Constrained Nanonodes: Context}
 
The usual energy lifecycle of an power-constrained nanonode is depicted in Figure~\ref{fig:intermittent}.
At certain points in time (i.e., the ``turn-off threshold'') the energy of the nanonode will be at a critically low level and the nanonode will turn off. 
At a certain later point in time, the nanonode will have harvested enough energy to turn on (i.e., the ``turn-on threshold'') and will become operational again.  
Intuitively, the nanonode will continue to harvest energy if it is turned on until its energy level reaches the maximum storage capacity, as depicted in the figure.
During transmission, reception, or any other operational periods (e.g., idling, sensing, actuation), the nanonode will loose energy, while simultaneously gaining some due to harvesting.

We consider nanoscale energy harvesters that exploit piezoelectric effect of ZnO nanowires~\cite{wang2008towards}, where the energy is harvested in nanowires' compress-and-release cycles.
The harvested energy can be specified with the duration of the harvesting cycle $t_{cycle}$ and the harvested charge per cycle $\Delta Q$.
Capacitor charging through energy harvesting can be accurately modeled as an exponential process~\cite{jornet2012joint}, accounting for the total capacitance $C_{cap}$ of the nanonode, where $C_{cap} = 2 E_{max} / V_g^2$, i.e., $C_{cap}$ depends on the maximum energy storage capacity $E_{max}$ and the generator voltage $V_g$.  
In the modeling, it is required to know in which harvesting cycle $n_{cycle}$ the nanonode is, given its current energy level $E_{n_{cycle}}$, which can be derived from~\cite{jornet2012joint} as follows:

\begin{equation}
\label{eq1}
n_{cycle} = \ceil*{\frac{- V_g C_{cap}}{\Delta Q} ln\left(1 - \sqrt{\frac{2 E_{n_{cycle}}}{C_{cap} V_g^2}}\right)}.
\end{equation}

The energy in the next energy cycle $n_{cycle} + 1$ is then: 

\begin{equation}
\label{eq2}
E_{n_{cycle+1}} = \frac{C_{cap} V_g^2}{2} \left(1- e^{-\frac{\Delta Q (n_{cycle} + 1)}{V_g C_{cap}}}\right)^2.
\end{equation}
\vspace{-3mm}

\subsection{Localizing Software-Defined Metamaterials}

\begin{figure}[!t]
\centering
\includegraphics[width=0.58\linewidth]{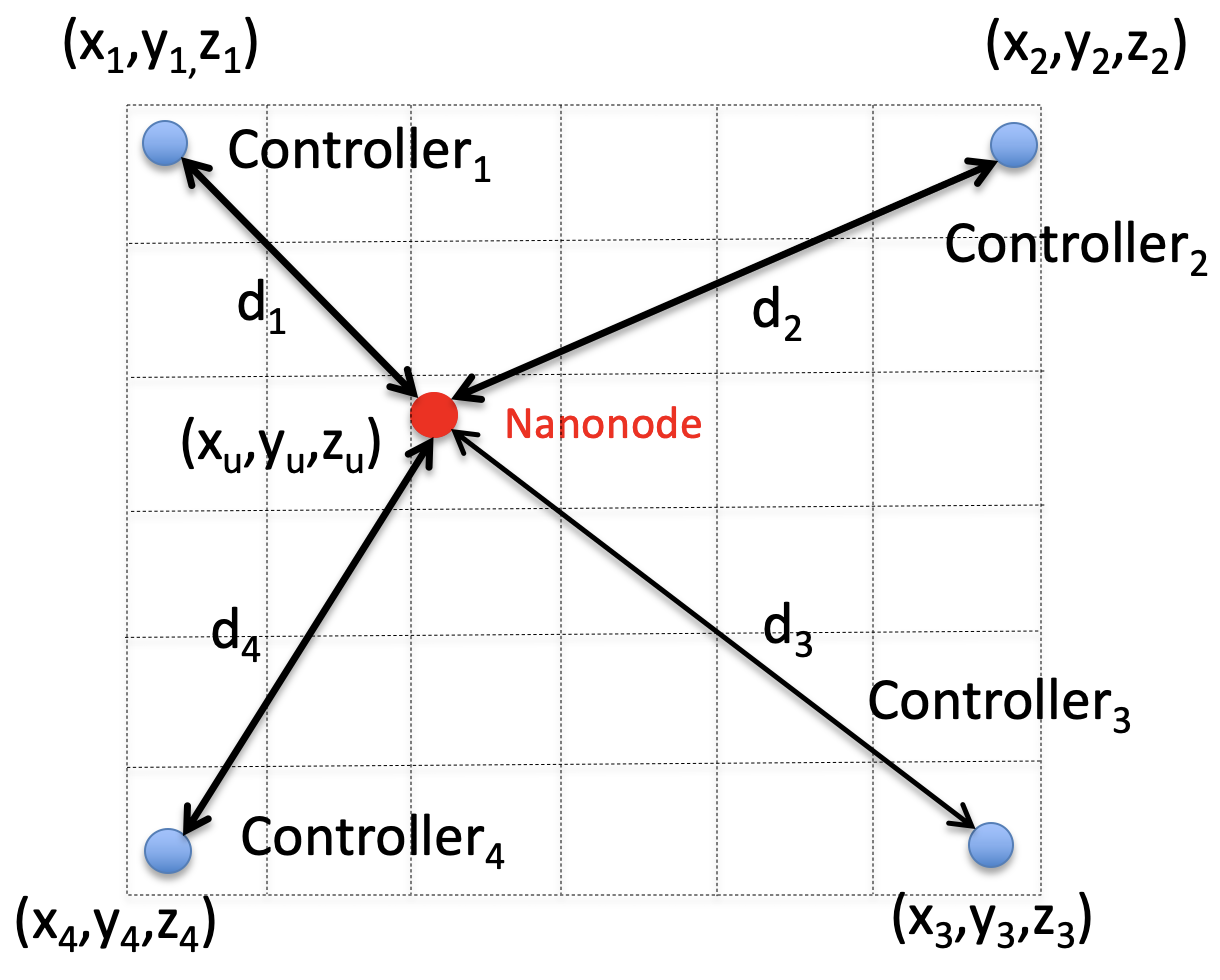}
\caption{Trilateration setup}
\label{fig:localization_setup}
\end{figure}

We envision the localization setup as depicted in Figure~\ref{fig:localization_setup}.
Specifically, there is a number of controllers (four depicted in the figure) serving as localization anchors.
They are not power-constrained and their locations are known with perfect accuracy. 
The assumption of the controllers not being power-constrained (i.e., mains- or battery-powered) has been widely used in the literature (e.g.,~\cite{liaskos2015design,lemic2019assessing}).
This is mainly because there are substantially less controllers than the metamaterial elements that they are controlling (i.e., one controller is envisioned to control hundreds of elements~\cite{lemic2019assessing}), hence their wiring requirements are less stringent.
The assumption that their locations are known is reasonable as their locations can be fixed and measured (e.g., when an SDM is mounted on a rough surface).
Adversely, their locations can be estimated by utilizing some of the traditional approaches, as they are not size-, resource-, or power-constrained.  
Localization could also be done in two-steps, where first the controllers are localized by utilizing two-way ToF-based trilateration, followed by localizing the nanonodes. 
This can be done as the controllers are not power-constrained, thus they can utilize higher transmit power for localization, which in turn would attribute to the enhanced range of localization.

Under the above assumptions, the localization process includes each controller transmitting one or multiple TS-OOK pulses, which are upon reception retransmitted by the nanonode whose location is to be estimated.
If the retransmitted signals are received by the controller, the distances between the controller and the nanonode can be estimated from the two-way ToF measurements.
The reasons why the retransmitted signals could be missed are two-fold. 
First, the nanonode's energy could be depleted, hence it would not be able to retransmit the original signal(s).
Second, due to mobility (e.g., stretching, bending, or rolling of an SDM) a controller could move out of the communicating range of the nanonode.   
If the distances between at least four controllers and the nanonode can be estimated, a standard trilateration approach can then be utilized for estimating the unknown location of the nanonode in a \ac{3D} space.  
More details on how to estimate the distance between each controller and the nanonode, as well as the basics of trilateration, can be found in e.g.,~\cite{karl2007protocols}.

\begin{figure}[!t]
\centering
\includegraphics[width=0.78\linewidth]{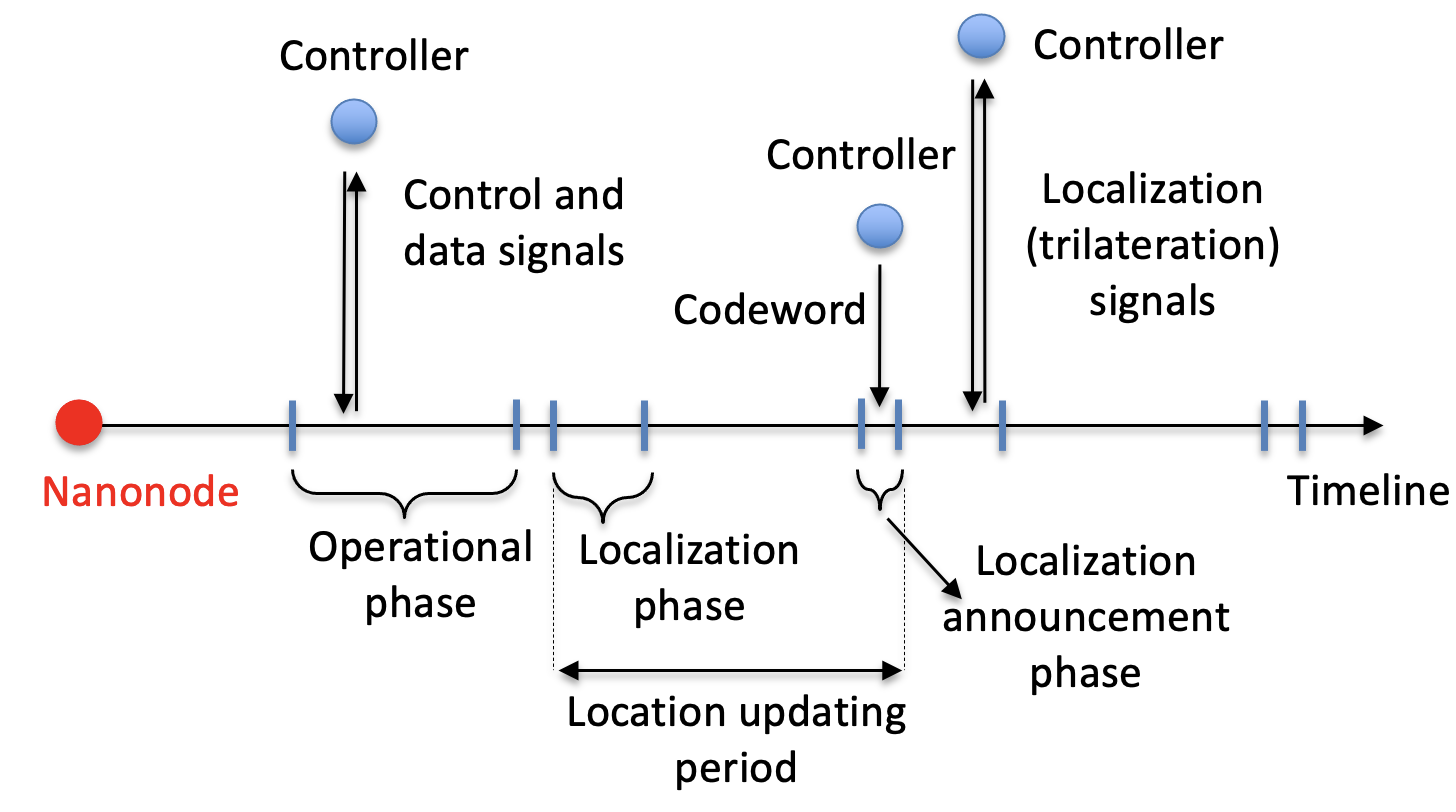}
\caption{Operational timeline of a nanonode}
\label{fig:operational_timeline}
\end{figure}

The operational timeline of the nanonode whose location is to be estimated is given in Figure~\ref{fig:operational_timeline}. 
Due to potentially continuous mobility, the location of the nanonode will usually have to be estimated periodically. 
Hence, we envision three phases in the operational timeline, i.e., the operational, localization announcement, and localization phase.
During the operational phase the nanonode performs its envisioned functionalities (e.g., sensing, actuation).
In the localization announcement phase, a distinct set of signals (i.e., a codeword) is sent by the controller(s) to the nanonodes to announce the beginning of the localization phase. 
This is done as the nanonodes internal clocks are not synchronized with the controllers, hence there is a need to announce a ``switch'' in their reactivity to the signals issued by the controllers, i.e., the end of the operational and the beginning of the localization phase.   
Note that this does not imply the need for synchronizing the nanonodes with the controllers, only the need for coarse-grained announcement of the switch between operations. 
In the localization phase, the locations of the nanonodes are envisioned to be estimated.   
The frequency of location estimation in this scenario is application-specific and depends on the location updating period, as shown in the figure. 

\section{Evaluation Methodology}
\label{methodology}

The main goal of the evaluation is to establish the accuracy and availability of the THz-based two-way \ac{ToF} approach for localization in SDMs.  
These performance metrics are to be derived as a function of a large number of relevant system parameters, i.e., operational bandwidth, energy harvesting rate, location update period, spacing between metamaterial elements, receiver sensitivity, number of TS-OOK pulses used for \ac{ToF} estimation, number and constellation of localization anchors, errors in the localization of the anchors, nanonodes' mobility patterns, nanonodes' energy consumption and harvesting patterns, and metamaterial-specific attenuation.  

The localization accuracy is characterized by the localization error, which is defined as the Euclidean distance between the true and estimated locations of a nanonode~\cite{moayeri2016perfloc,moayeri2018perfloc}.  
The localization availability for one attempt at localizing a nanonode is defined as a binary value indicating the success of localizing the nanonode or the lack thereof. 
Each instance of the results depicted in Figures~\ref{fig:harvesting} to~\ref{fig:mobility} provides insights summarized over all nanonodes and iterations of the simulation (cf., Table~\ref{tab:paramaters}).
The localization accuracy in each instance of the results is depicted in a regular boxplot fashion.
In an individual boxplot of the localization errors, the boxes extend from the lower to upper quartile of the derived localization errors, with an orange line at the median. 
The whiskers extend from the boxes to show the range of the localization errors, while the flier points represent the outliers in the localization errors, with the outliers defined as the lowest and highest 5 percentiles of the errors.
Moreover, the localization availability has been depicted as a ratio between the number of successful and the total number of localization attempts, summarized over all nanonodes and iterations of the simulation.

In addition, we aim at deriving the first rule-of-thumb insights on the latency of estimating location information, which is among the most important parameters for characterizing the performance of localization solutions~\cite{van2013evarilos}, however currently lacking in the context of SDM localization. 
Localization latency in this context essentially equals the duration of the localization phase (cf., Figure~\ref{fig:operational_timeline}) and consists of the time needed to obtain the ToF characterization for all controller-nanonode pairs plus the time needed to estimate the nanonodes' locations based on the derived ToFs.  
The summary of the default simulation parameters is given in Table~\ref{tab:paramaters}. 

\begin{table}[!t]
\small
\begin{center}
\caption{Simulation parameters}
\label{tab:paramaters}
\begin{tabular}{l r}
\hline
\textbf{Parameter} & \textbf{Value} \\
\hline
Number of nanonodes and controllers & (25x25) 625 \\
Distance between nanonodes [mm] & 9.0 \\
Generator voltage $V_g$ [V] & 0.42 \\
Energy consumed in pulse reception $E_{R_{X pulse}}$ [pJ] & 0.1 \\
Energy consumed in pulse transmission $E_{T_{X pulse}}$ [pJ] & 1.0 \\
Maximum energy storage capacity [pJ] & 800 \\
Turn OFF/ON thresholds [pJ] & 10/0 \\
Harvesting cycle duration [ms] & 20 \\
Harvested charge per cycle [pC] & 6 \\
Transmit power $P_{T_X}$ [dBm] & -20 \\
Data packet size [bits] & 8 \\
Operational bandwidth [THz] & 1 \\ 
Receiver sensitivity [dBm] & -110 \\ 
Operational frequency [THz] & 1 \\ 
Simulation time [\# of iterations] & 1000 \\
Localization update period [sec] & 0.1 \\
Environment-specific attenuation [dB] & 0 \\ 
\hline
\end{tabular}
\end{center}
\vspace{-5mm}
\end{table}

We have carried out our evaluation by means of simulation, with the primary reason for not aiming at a full scale experimental evaluation coming from the fact that there are currently no operational prototypes of THz-operating power-constrained SDMs available for experimentation.  
To carry out the evaluation, we have developed a Python-based simulator. 
We have done that, in contrast to using some (mostly ns-3 based) of the established and readily available simulators (e.g., TeraSim~\cite{hossain2018terasim}) for evaluating the performance of THz-operating nanonetworks, as such simulators feature a fairly heterogeneous simulation capabilities, resulting in much worse execution time and scalability for the problem at hand.
Utilization of the in-house Python simulator allowed us to increase the number of considered system parameters and derive a number of insights in a reasonable time-frame. 
Our simulator consists of four main modules: channel modeling, two-way ToF-based trilateration, nanonode's energy consumption and harvesting, and performance metrics calculation). 
The output of the channel modeling module of our simulator has been compared to the same output of TeraSim~\cite{hossain2018terasim} and are comparable, suggesting that our implementation of the module is correct. 
The fact that the other three main modules would have to be implemented regardless of the simulator makes us confident that our simulator is able to derive the same insights for \textit{the problem at hand} as TeraSim. 

In the simulator, we have initially defined a set of 625 (25x25) nanonodes and controllers in a grid-like fashion. If not stated differently, the four controllers are positioned in the corners of the grid.
The default distance between neighboring nanonodes was set to 9.0~mm, which correlates to spacing needed for controlling electromagnetic waves in the mmWave frequencies (i.e., with the operational frequency of 25~GHz, assuming the spacing equals to 3/4 of the wavelength).
Hence, they were not power-constrained and (if not stated differently) their locations were known and fixed.      
The other nanonodes are assumed to be powered through energy harvesting, with their energy levels modeled using Equations~\ref{eq1} and~\ref{eq2}. 
Moreover, we assumed that the energy is harvested from two different environmental sources, specifically air-vibrations and RF-based power transfer.
We have selected these two harvesting options with the following intuitions. The air-vibrations represent an ubiquitous energy harvesting source that can be utilized in a large variety of deployment environments and conditions. The RF-based power transfer using metamaterial elements themselves~\cite{amer2020comprehensive} is tightly related to one of the most prominent applications of SDMs, namely the intelligent wireless environments. 
In the intelligent wireless environments, the SDMs will be used for controlling the properties of RF waves, hence it is natural that the energy of these waves can also be utilized for powering the power-constrained nanonodes~\cite{di2019smart}.  
In agreement with the existing literature, we have specified the default harvesting cycle duration of 20 and 1.71~ms for air-vibrations~\cite{jornet2012joint} and RF power transfer~\cite{canovas2018nature}, respectively.
Moreover,~\cite{jornet2012joint,canovas2018nature} report the average harvesting charge $\Delta Q$ of 6~pC for one ZnO nanowires' compress-and-release cycle. 
As commonly done in the literature (e.g., in TeraSim~\cite{hossain2018terasim}), we model these charges using Gaussian distributions with 6~pC as the mean value and 0.6 pC as the standard deviation.
We consider a capacitor with the energy storage capacity of 800~pJ, which is in correspondence to the existing literature~\cite{jornet2012joint}.
The initial energy levels of the nanonodes have been drawn from a uniform distribution bound by the energy storage capacity. 

In order to establish the availability of the localization service, we have also considered the energy required for the necessary signaling between the controllers and nanonodes whose location is to be estimated.
In TS-OOK, the information is transmitted by short time-spread pulses or silences, where the vast majority of energy is spent in their transmission and reception, as originally shown by Jornet and Akyildiz~\cite{jornet2012joint}, which is also the assumption we utilize in this work.    
Specifically, the energies consumed in transmitting ($E_{T_{X pulse}}$) and receiving ($E_{R_{X pulse}}$) a TS-OOK pulse are set to 1 and 0.1~pJ~\cite{jornet2012joint,hossain2018terasim}, respectively.
Furthermore, the transmit power is set to $-20$~dBm, which is again in line with the existing literature~\cite{jornet2012joint,hossain2018terasim}. 
The energy is assumed to be consumed during the operational, localization announcement, and localization phases of a nanonode. 
During the localization phase, the energy can be consumed in the reception and retransmission of a TS-OOK pulse (or multiple TS-OOK pulses) for deriving the two-way ToF.
In the operational phase, the energy is, if not stated otherwise, assumed to be consumed in the reception of an 8 bits long packet, with the bits (i.e., logical ``0''s and ``1''s) being drawn from a uniform distribution.    
The reception of such a packet mimics an instruction sent to a nanonode for which an 8 bits long packet is usually sufficient~\cite{lemic2019survey,abadal2020programmable}.  
Note that we also consider various other relevant energy consumption patterns such as the nanonodes transmitting packets, performing sensing and actuation, and various combinations of these operations, as in more details discussed in the following section. 
Acknowledging that other types of sequences are possible, we assume the localization announcement is performed by transmitting a sequence of three TS-OOK pulses.
Hence, we model the nanonode's energy consumption in the localization announcement phase with the energy consumed in reception of three TS-OOK pulses.    
The default operational bandwidth is set to 1~THz, equaling the theoretical limit derived in~\cite{abadal2019graphene}, in order to demonstrate the theoretically achievable localization accuracy in SDMs.  
Note that, although 1~THz of bandwidth (implying comparable sampling frequency at the nanonode level) is not practically feasible, different approaches exist for two-way ToF sampling that could be used for mitigating this limitation and effectively result in two-way ToF ranging errors that could be achieved for the THz-frequent sampling, as in more details discussed in Section~\ref{open_challenges}.

If not stated otherwise, from the initial grid-like distribution the nanonodes are ``moved'' randomly in the area of sizes $(x,y,z)=(d,d,d/2)$, with $d$ being the distance between two controllers on the same edge.
This random selection of locations inside a bounded area resembles a scenario in which a flexible SDM is attached to an uneven convex surface, such as in the smart textiles~\cite{tian2019wireless}.
In addition, we consider the mobility of the nanonodes resembling a half-spherical and half-cylindrical patterns bound by the same area $(x,y,z)=(d,d,d/2)$ (Figure~\ref{fig:mobility}).   
These patterns are relevant as they mimic the mobility of SDMs bent around different types of elastic structures for e.g., structural health monitoring~\cite{ozbey2014wireless} or mitigating the exposure to electromagnetic radiation~\cite{gasmelseed2014effects}.

Two-way ToF measurements are derived from the true distances between each nanonode and each controller, to which bandwidth-dependent zero-mean Gaussian-drawn variabilities are added.   
The standard deviation of the variability is derived as the ratio between the speed of light and bandwidth (i.e., the raw resolution), which is an often utilized method for simulating the ToF variability~\cite{wirstrom2015localization,lemic2016localization}.

Location estimates can be generated if the two-way ToF measurements between a given nanonode and all 4 controllers are obtained.
This is characterized by comparing the strength of the received signal with the receiver sensitivity, i.e., if the received signal strength is higher than the receiver sensitivity, the signal is considered as received.
The received signal strength $P_{Rx}$ is obtained by subtracting frequency-selective absorption and spreading losses from the transmit power $P_{Tx}$, which is a THz nanoscale channel modeling method often used in the literature (e.g.,~\cite{jornet2011channel,kokkoniemi2014frequency}).
The received signal strength is given as follows, with $d$, $f$, and $c$ being respectively the distance between devices, operating frequency, and the speed of light. Moreover, $k(f)$ is a frequency-dependent medium absorption coefficient, with its values obtained from the HITRAN database~\cite{rothman1987hitran}:

\begin{equation}
\label{eq3}
P_{Rx} [dBm] = P_{Tx} - k(f) d 10 log_{10}(e) - 20 log\left(\frac{4 \pi f d}{c}\right).
\end{equation}

\begin{figure}[!t]
\centering
\includegraphics[width=\linewidth]{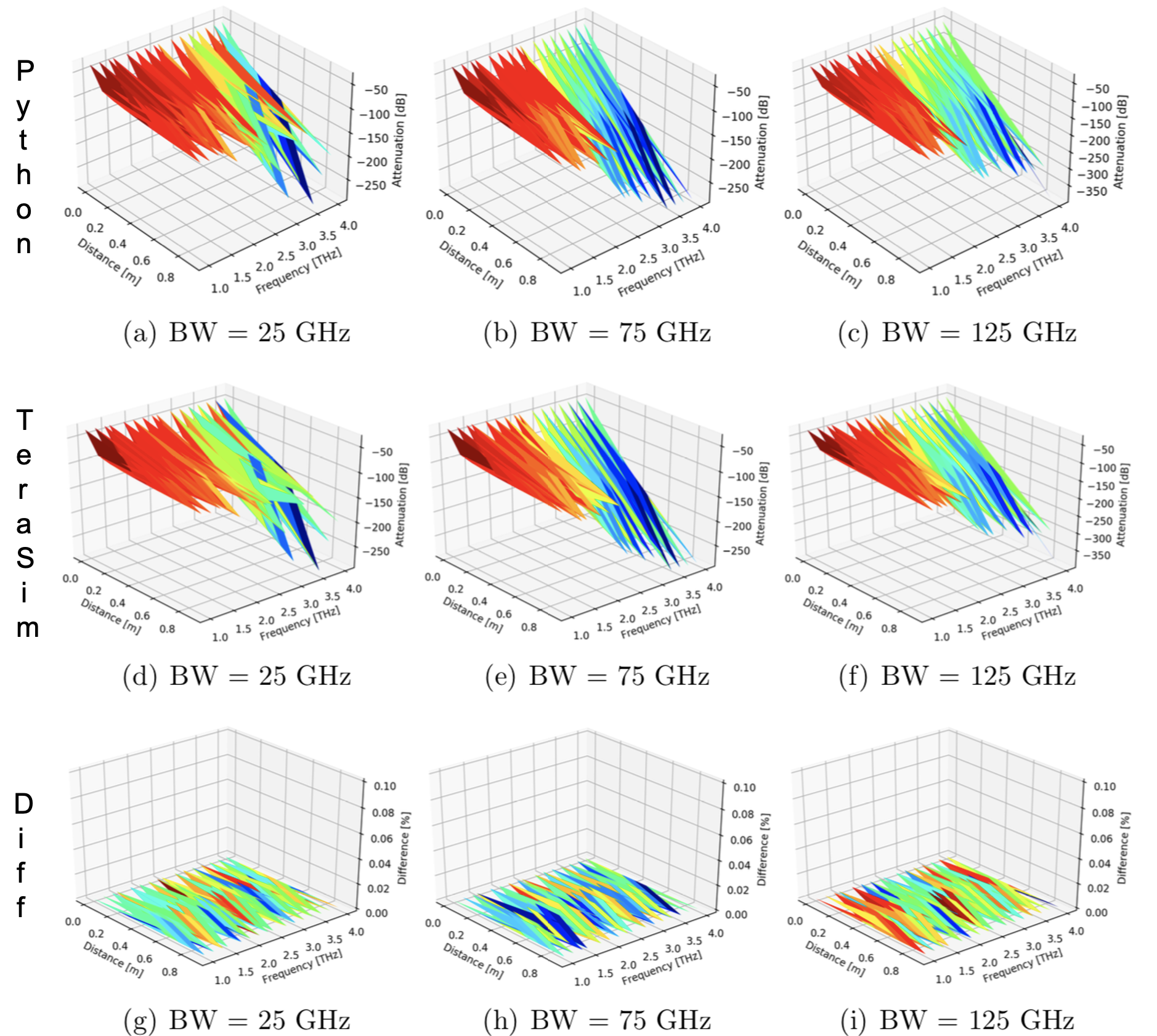}
\caption{Attenuation (absorption and spreading losses) as a function of operating frequency, bandwidth, and distance between the devices}
\label{fig:channel}
\end{figure}

The modeled attenuation as a function of operating frequency, bandwidth, and distance between the devices is given in Figure~\ref{fig:channel}.
As visible from the figure, the attenuation increases with the increase in the operating frequency and distance between the devices, which is an intuitively expected behaviour.
Moreover, the utilized bandwidth affects the attenuation due to the frequency selectivity in the propagation channel, which is again an expected behaviour aligned with propagation modeling in TeraSim, as visible from the figure.
Finally, the error of our implemented propagation model is well below 1\% compared to TeraSim in all cases. 
We hypothesize this small error to be due to differences in rounding and precision between the two used programming languages (i.e., Python and C++).	

Given that the flexible SDMs will in the majority of cases be ``attached'' to a certain surface, we also consider an environment-specific attenuation contribution $A_{ENV}$ to the received signal strength.
With the environment-specific attenuation contribution $A_{ENV}$, we are also abstracting various imperfections that are certainly to be expected in reality, but cannot be fully captured using a simulation-based approach.
Some examples of abstracted imperfections include the antenna coupling effects, imperfections in the omnidirectional radiation, potential interferences, etc.
Given that this attenuation contribution is hardly quantifiable based on the existing literature and will intuitively depend on the utilized hardware (e.g., antenna coupling efficiency) and the deployment environments (e.g., type of surface to which the SDM is attached), we characterize the effect of an increasing environment-specific attenuation contribution $A_{ENV}$ on the performance of the proposed localization system.
We consider such modeling to be appropriate, as it encapsulates different types of signal degradations (e.g., scattering,  absorption) and establishes their effects, thus also the effects of different surfaces, on the performance of the proposed system. 
The received signal strength is finally characterized as: 

\begin{equation}
\label{eq4}
P_{Rx} [dBm] = P_{Tx} - k(f) d 10 log_{10}(e) - 20 log\left(\frac{4 \pi f d}{c}\right) - A_{ENV}.
\end{equation}


\section{Evaluation Results}
\label{results}

\subsection{Localization Accuracy and Availability}

\begin{figure}[!t]
\centering
\includegraphics[width=0.75\linewidth]{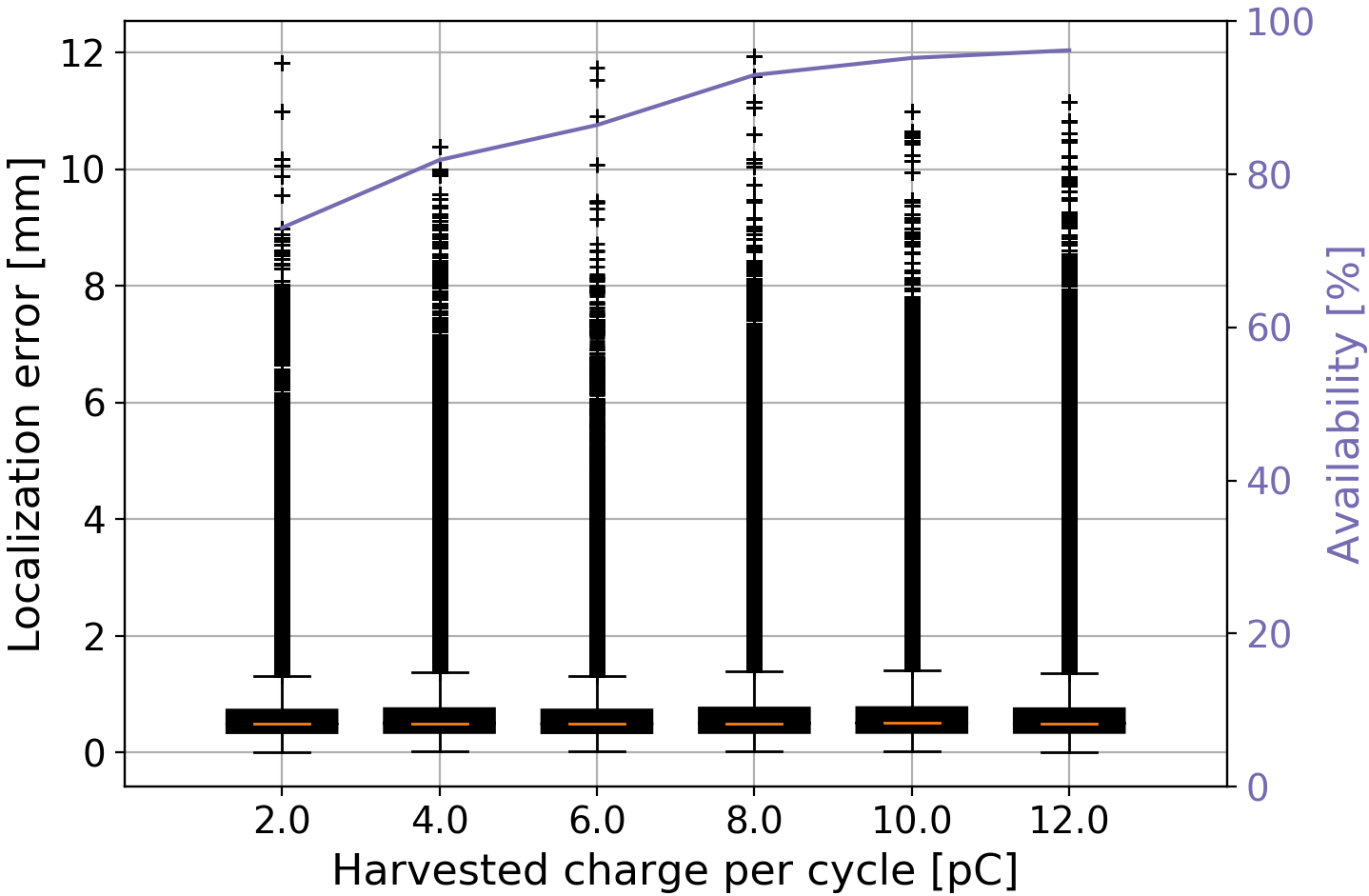}
\caption{Accuracy and availability vs. energy harvesting rate}
\label{fig:harvesting}
\end{figure}

\begin{figure}[!t]
\centering
\includegraphics[width=0.75\linewidth]{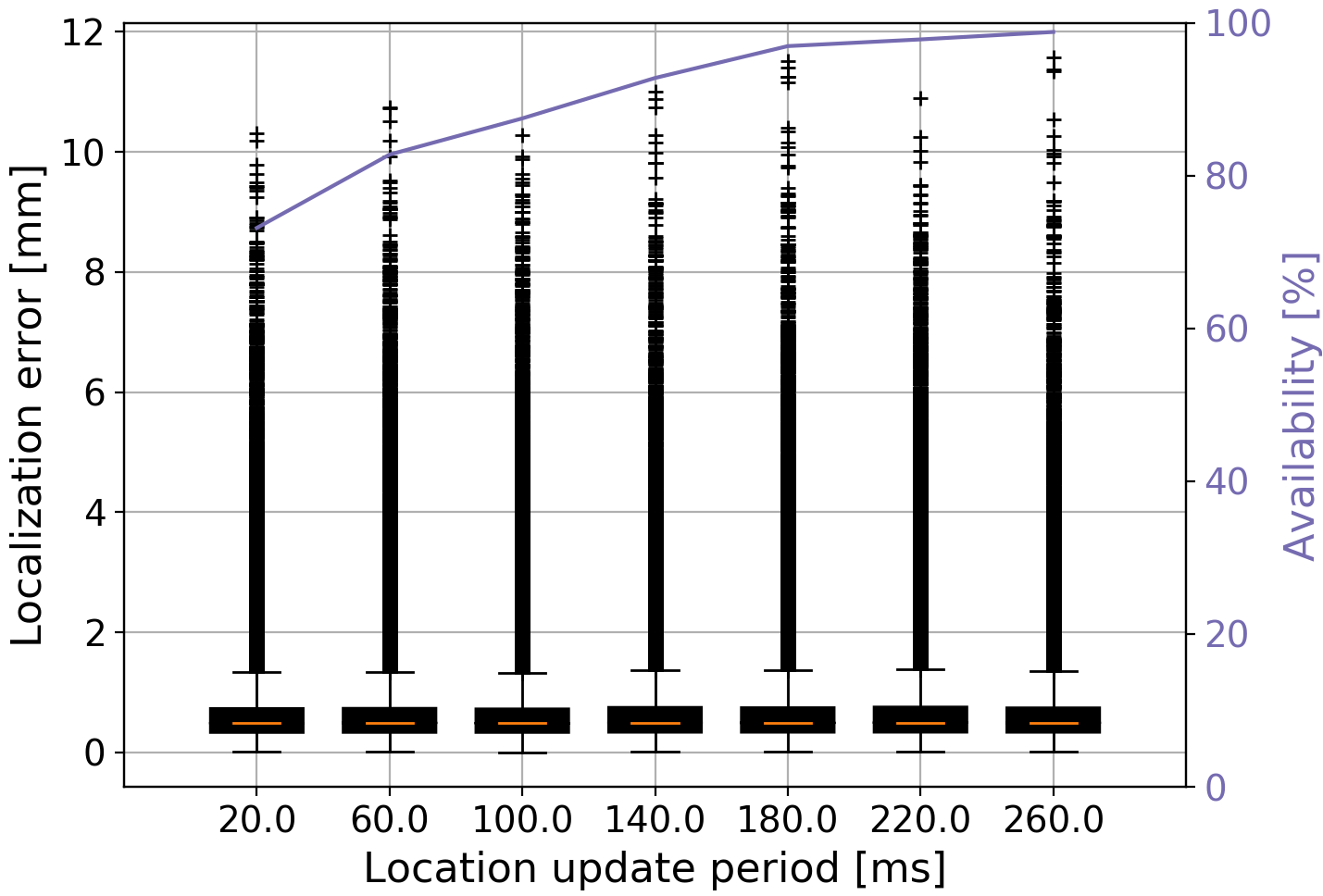}
\caption{Accuracy and availability vs. location update period}
\label{fig:loc_frequency}
\end{figure}

Figures~\ref{fig:harvesting} and~\ref{fig:loc_frequency} depict the accuracy and availability of the localization service as a function of the energy harvesting rate and location update period, respectively.
As depicted in the figures, the energy harvesting rate and location update period do not significantly affect the localization accuracy.
This is because all the nanonodes are in the default setup in the range of all controllers, regardless of the energy harvesting rate and location update period.
Hence, these parameters only affect the probability of generating two-way ToF measurements, and not their quality.
In other words, if the measurements are obtained, location can be estimated with a consistent accuracy.  
The localization service achieves less than 1~mm in terms of average localization error, which demonstrates the feasibility of the two-way ToF-based trilateration in SDMs. 
Moreover, the outliers (95$^{th}$ percentile of all estimates) experience localization errors of less than 12.0~mm, which again shows the promise of the two-way ToF-based trilateration service.

For the parameters reported in Table~\ref{tab:paramaters}, the availability of the service is around 90\%, with the only reason for not being able to estimate the locations resulting from the depletion of the available energy. 
This again demonstrates the potential of the proposed system for localization in SDMs. 
As shown in the figures, the availability of the localization service is enhanced as the energy harvesting rate or location update period increase.
For example, if the harvested charge is increased from 2 to 10~pJ per cycle, the availability is increased by roughly 10\%.
Similarly, if the localization update period increases from 20 to 260~ms, the availability increases from 80 to 100\%.
The reason for that is the fact that higher harvesting rates represent more harvested energy, while less frequent location updates yield less energy consumed over time.
Therefore, the nanonodes more often have sufficient energy to generate two-way ToF measurements and consequently estimate the location, which enhances the availability of the localization service.   
Though the results depicted in Figures~\ref{fig:harvesting} and~\ref{fig:loc_frequency} are intuitive, we present them to serve as an argument demonstrating the correct implementation of our simulation setup.

\begin{figure}[!t]
\centering
\includegraphics[width=0.75\linewidth]{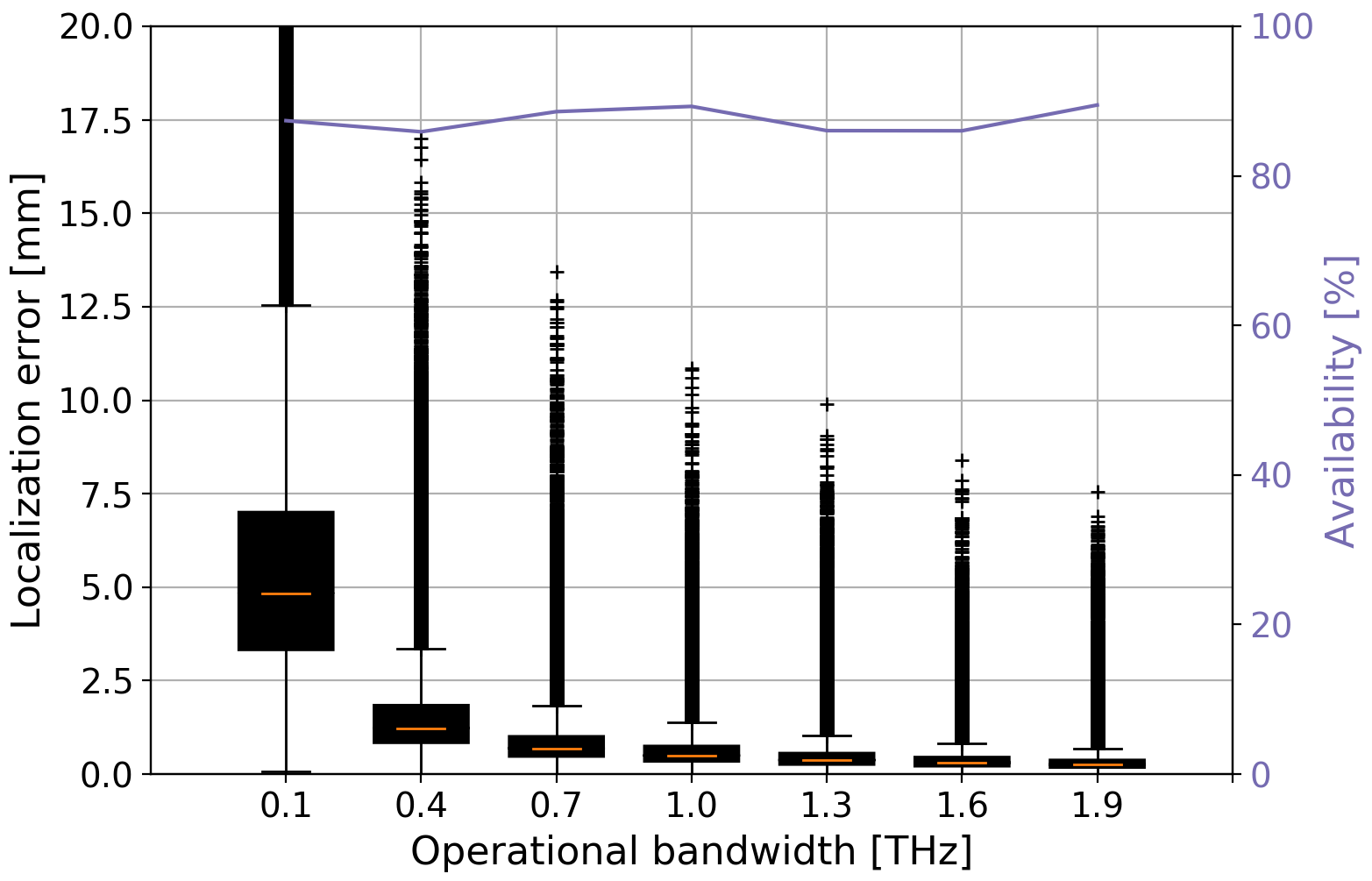}
\caption{Accuracy and availability vs. operational bandwidth}
\label{fig:bandwidth}
\end{figure}

\begin{figure}[!t]
\centering{}
\includegraphics[width=0.75\linewidth]{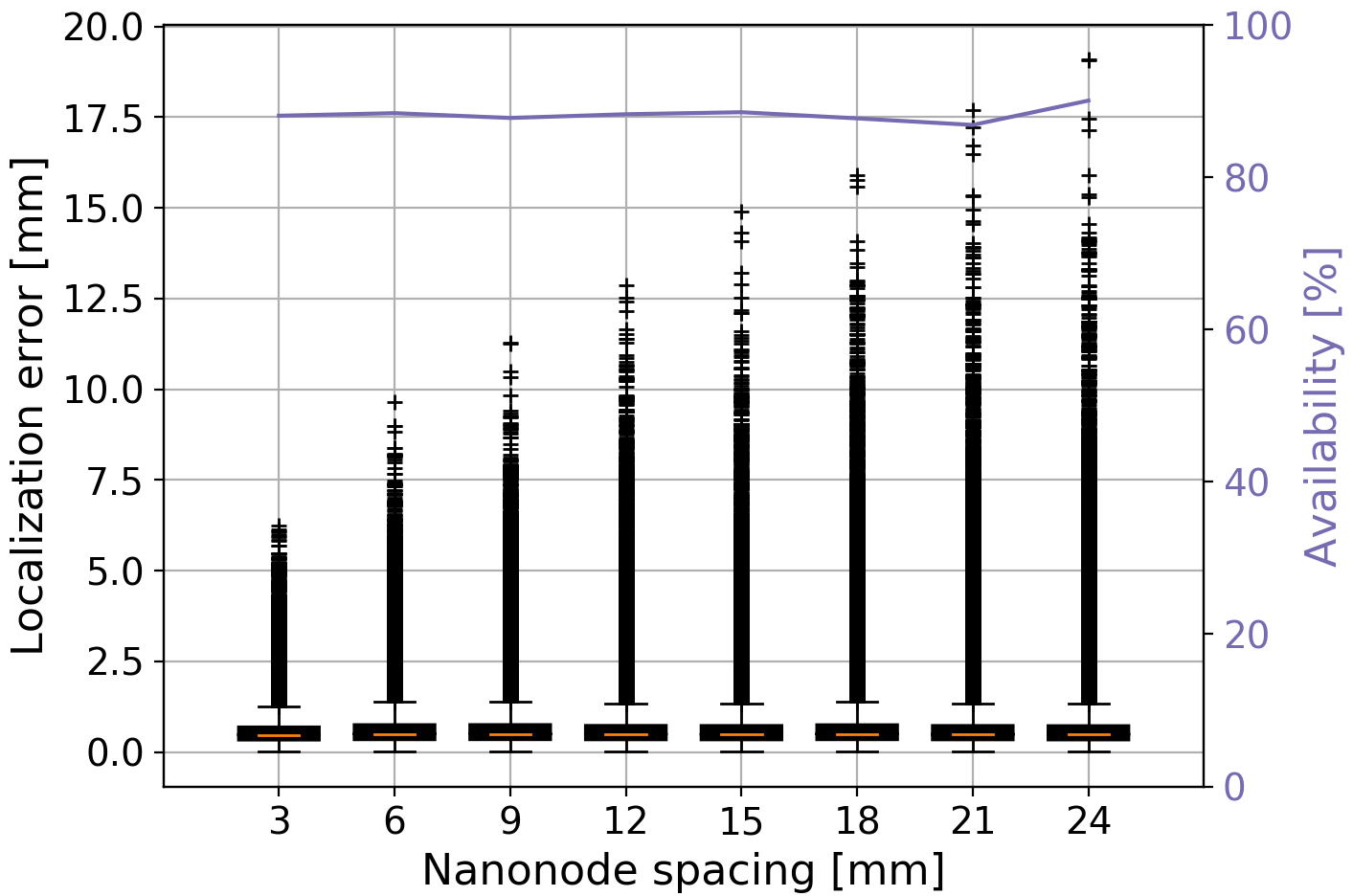}
\caption{Accuracy and availability vs. spacing between nanonodes}
\label{fig:spacing}
\end{figure}

Figures~\ref{fig:bandwidth} and~\ref{fig:spacing} depict the localization accuracy and availability in relation to the utilized bandwidth and spacing between nanonodes.  
As shown in the figures, both parameters have an impact on the localization accuracy, while not influencing the availability of the service.
This is because the increase in the utilized bandwidth increases the achievable signal sampling rate.
This in turn reduces the variability of the two-way ToF measurements, benefiting the accuracy of localization.
For example, if the utilized bandwidth is increased from 0.1 to 1~THz, the average localization error is reduced from roughly 5 to less than 1~mm.   
Similarly, the spacing between the nanonodes affects the localization accuracy, however in this case only the outliers are affected.
This is because as the spacing increases, the sizes of the localization area increase as well, which in turn results in larger errors of the outliers.
This is a well-known behavior in more traditional localization approaches, e.g.,~\cite{behboodi2017hypothesis,lemic2016toward}. 

We argue that the results presented in Figure~\ref{fig:bandwidth} demonstrate that the proposed localization approach can provide solid performance even with constrained bandwidth (in the THz context), which could be a beneficial feature for avoiding interference in case of multiple closely-located SDMs by utilizing non-overlapping frequency bands.   
Note that it is well-established that RF-interference (even only data transmission in the same frequency band) can significantly degrade the performance of RF-based localization solutions~\cite{lemic2015experimental,behboodi2015interference}.
Moreover, based on the results shown in Figure~\ref{fig:spacing}, we argue that the proposed localization solution can operate well across different frequencies of the electromagnetic waves that are being manipulated by the SDM. 
Specifically, assuming that the spacing between metamaterial elements equals 3/4 of the wavelength, we show that localization is feasible for SDMs used for controlling electromagnetic waves with frequencies ranging from cc. 9.4 (i.e., spacing of 24~mm) to 75~GHz (i.e., spacing of 3~mm), thus encapsulating the most interesting (at least currently) mmWave frequencies. 

\begin{figure}[!t]
\centering
\includegraphics[width=0.72\linewidth]{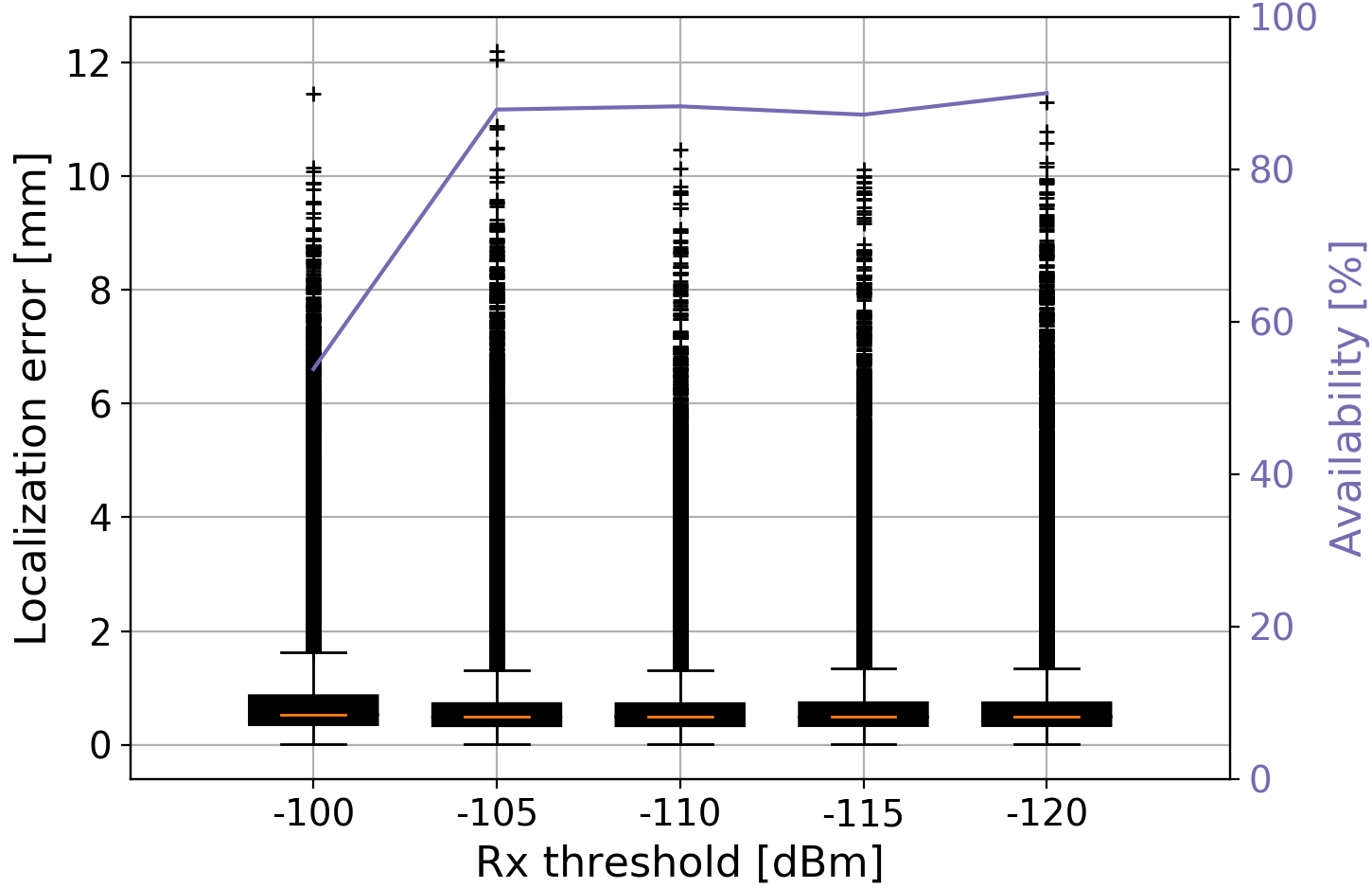}
\caption{Accuracy and availability vs. receiver sensitivity}
\label{fig:rx_threshold}
\end{figure}

\begin{figure}[!t]
\centering
\subfigure[3~mm spacing between neighboring nanonodes]{\vspace{-4mm}\includegraphics[width=0.7\linewidth]{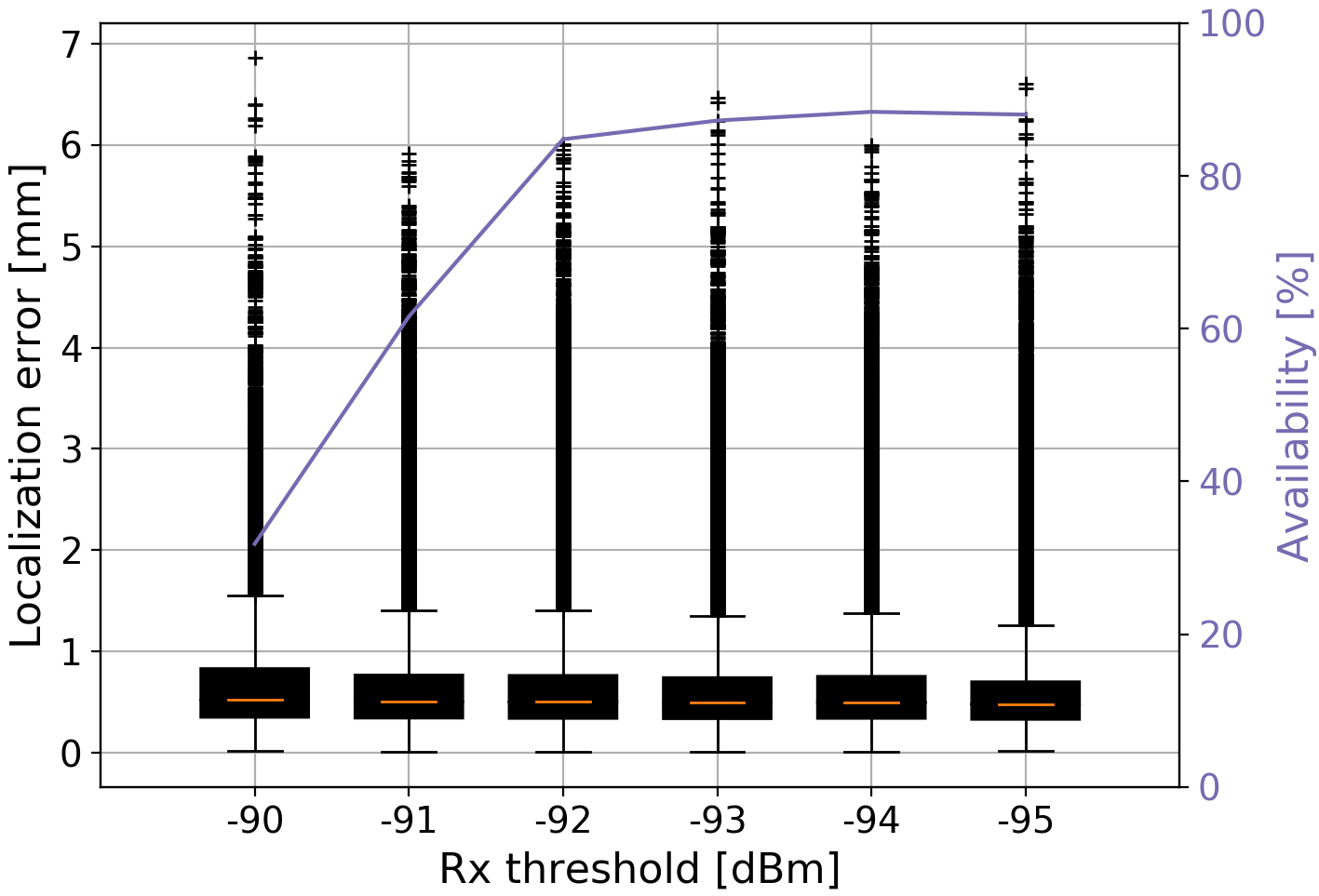}}
\subfigure[18~mm spacing between neighboring nanonodes\vspace{-4mm}]{\includegraphics[width=0.7\linewidth]{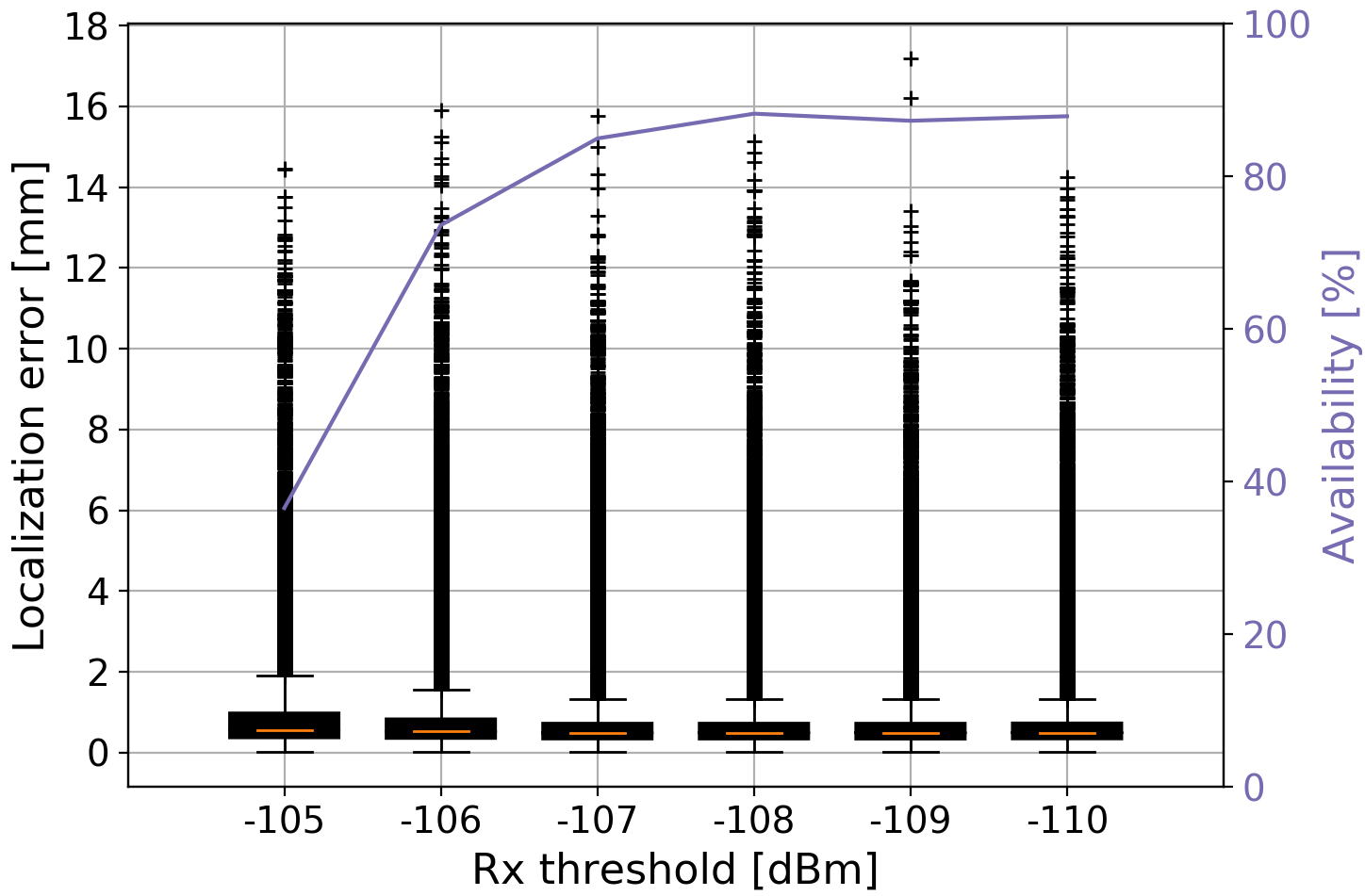}}
\caption{Accuracy and availability vs. receiver sensitivity for different spacings between nanonodes}
\label{fig:rx_sensitivity}
\end{figure}

As depicted in Figure~\ref{fig:rx_threshold}, the receiver sensitivity does not significantly affect the localization accuracy and availability in the default setup for the sensitivity levels below roughly -105~dBm.
This is because all the nanonodes are in the range of the controllers and experience relatively high received powers.
The effect on the accuracy and availability becomes more pronounced when the receiver sensitivity is low (i.e., above -105~dBm) or when the spacing between nanonodes further increases.
This is depicted in Figure~\ref{fig:rx_sensitivity}, where we show the interplay between spacing and receiver sensitivity for the spacings of 3~mm and 18~mm, while Figure~\ref{fig:rx_threshold} depicts the same for the spacing of 9~mm.
As visible in the figures, the availability of the localization service experiences a rapid drop if the receiver sensitivity is reduced below a certain value. 
This value, however, depends on the spacing between nanonodes, hence for relatively small spacings this drop is experienced for very low receiver sensitivities (e.g., around -92~dBm for the spacing of 3~mm, Figure~\ref{fig:rx_sensitivity}a) not expected in practice.   
As the spacings increase, the receiver sensitivity starts playing an important role, i.e., the drops in availability are experienced at the receiver sensitivity levels expected (cf., Figures~\ref{fig:rx_threshold}) or even unfeasible in practice (cf., Figure~\ref{fig:rx_sensitivity}b).
We believe a careful design of the localization system, accounting for the receiver sensitivity and spacings between metamaterial elements, will be needed for guaranteeing high availability of localization.
In case the spacings are too large and the availability of localization is not satisfiable, as is the case in Figure~\ref{fig:rx_sensitivity}b), introducing additional localization anchors will be required, which is a well-known approach in increasing the accuracy and availability of more traditional RF-based localization services~\cite{behboodi2017hypothesis}.

\begin{figure}[!t]
\centering
\includegraphics[width=0.75\linewidth]{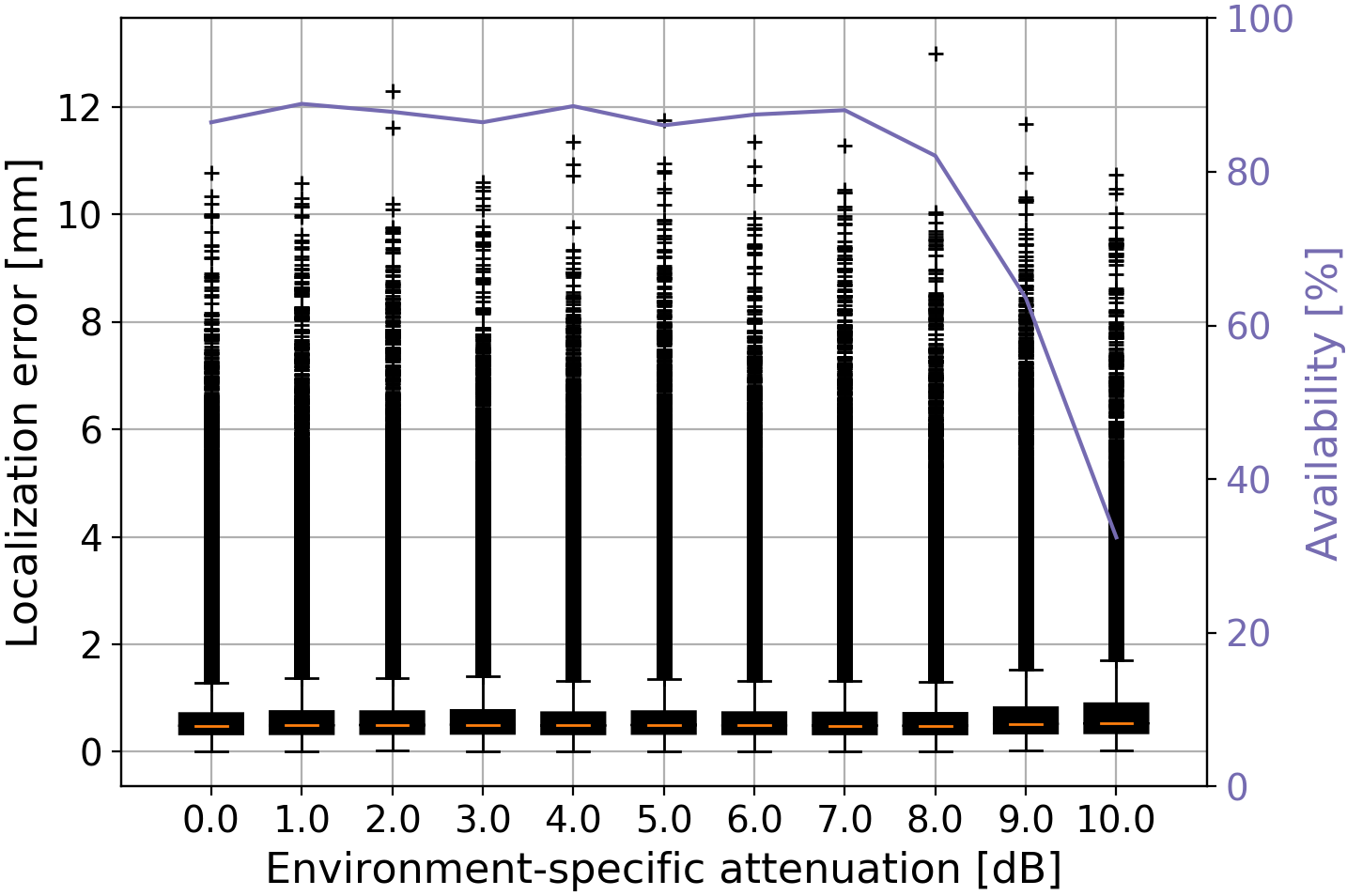}
\caption{Accuracy and availability vs. deployment-specific attenuation}
\label{fig:attenuation}
\end{figure}

\begin{figure}[!t]
\centering
\includegraphics[width=0.72\linewidth]{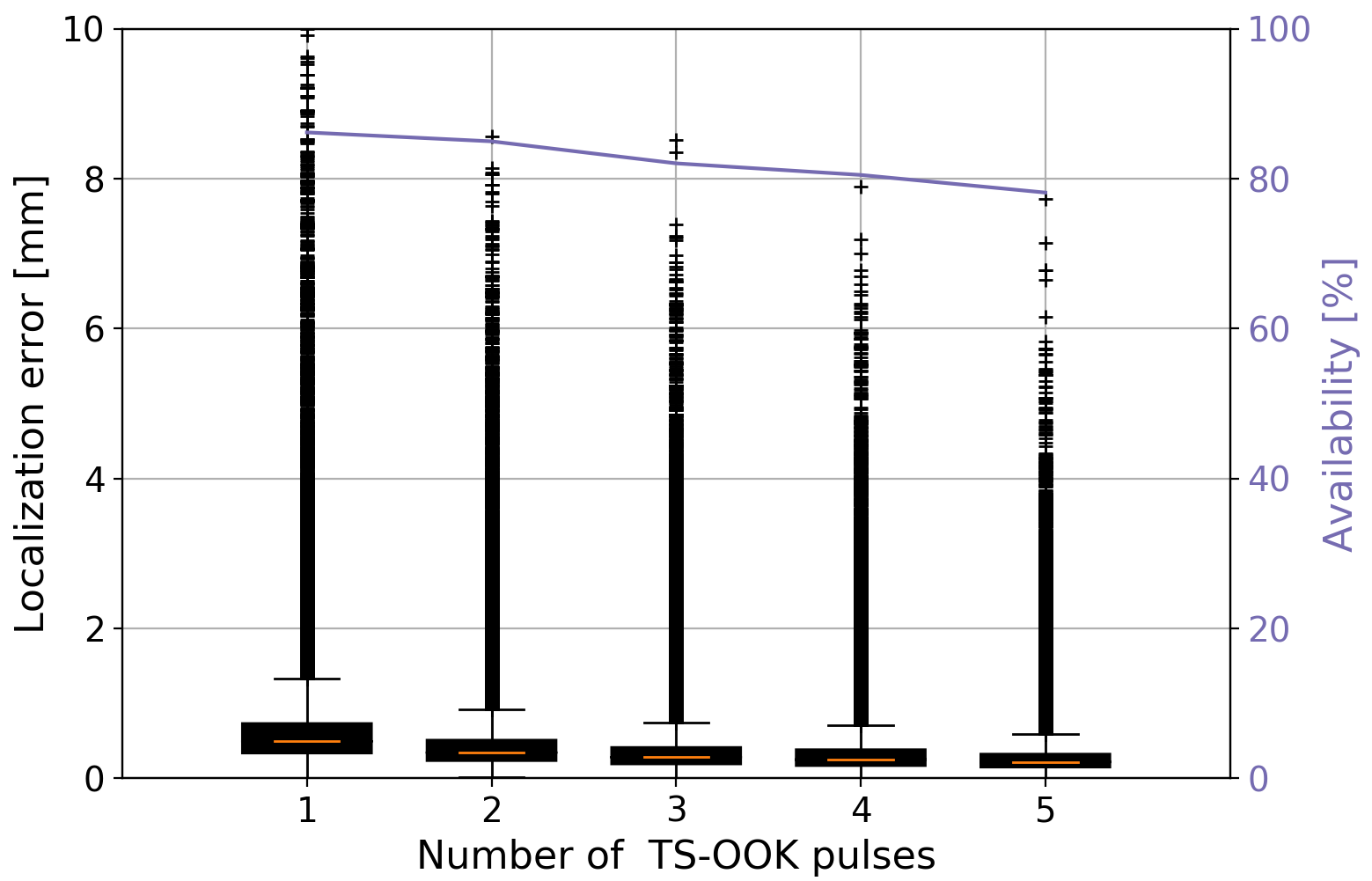}
\caption{Accuracy and availability vs. number of TS-OOK pulses used for two-way ToF derivation}
\label{fig:nr_pulses}
\end{figure}

The derivations presented above assume that the signal propagation in THz frequencies is not affected by the attenuation from a surface to which an SDM is mounted (i.e., Equation~\ref{eq3}), which is not a realistic assumption for some of the envisioned applications. 
Hence, in Figure~\ref{fig:attenuation} we also consider the environmental attenuation $A_{ENV}$ as a high-level approximation of all absorption and scattering losses from the surface on which an SDM is mounted (i.e., Equation~\ref{eq4}). 
As depicted in the figure, for the default simulation parameters from Table~\ref{tab:paramaters}, there is no significant effect of the environmental attenuation $A_{ENV}$ until roughly 8~dB when the availability of the localization service drops rapidly. 
Similar as before, the effect is only visible in case a nanonode is not in the range of some of the controllers, hence two-way ToF trilateration cannot be performed. 
In this case, additional localization anchors will have to be introduced in the system to cope with strong attenuation from the surface on which an SDM is mounted.
In more general terms, in the design of a system for localization in SDMs, one will have to account for the scattering and absorption losses from a particular surface, and the design of such a system will have to be environment specific for attenuation-heavy environments.
Nevertheless, initial studies in the literature report around 5~dB scattering and absorption losses for different types of sands~\cite{du2017characterisation}, and less than 5.6~dB for losses in ordinary clothing materials~\cite{dietlein2008absorption}, suggesting that for some of the potential use-cases the environment-specific designs might not be needed.

In Figure~\ref{fig:nr_pulses}, we depict the localization accuracy and availability as a function of the number of TS-OOK pulses used for two-way ToF estimation. 
As mentioned before, each controller is envisioned to transmit a TS-OOK pulse followed by its retransmission by a particular nanonode, which is then used for estimating two-way ToF for that controller/nanonode pair.
The same procedure can be repeated multiple times (x-axis in Figure~\ref{fig:nr_pulses}) for obtaining a number of two-way ToF estimates for a given controller/nanonode pair.
In the consequent step, these estimates can be averaged into a single two-way ToF estimate with lower expected variability, which in turn increases the localization accuracy, as depicted in the figure. 
Example-wise, the utilization of three averaged TS-OOK pulses instead of one reduces the maximum localization errors in the outliers from roughly 12 to 8~mm. 
As shown in the figure, the effect of increasing the number of TS-OOK pulses on the average localization accuracy is minor, while the latency of producing location estimates experiences a significant increase (more details in Section~\ref{latency}).
In addition and perhaps counterintuitively, the availability of localization slightly decreases with the increase in the number of TS-OOK pulses, as show in the figure.
This is because of the increase in the number of pulses results in higher energy consumption of the nanonodes (i.e., multiple transmissions and receptions needed for performing two-way ToF estimation), in turn increasing the chances of energy depletion. 
These observations have two important implications.
First, in the design of a localization solution for SDMs, one should consider the requirements that the application(s) to be supported pose on the underlying localization service.
Only if the localization accuracy with a single TS-OOK pulse cannot meet the requirements from the applications, one should aim at increasing the number of pulses, as this increase will inevitably affect the availability and latency of providing location estimates. 
Second, the above discussion amplifies the importance of considering multiple evaluation metrics for characterizing the performance of localization in SDMs.
In this particular case, the interplay between localization accuracy and latency of providing location estimates has been emphasized, although additional metrics could be considered, as in more details discussed in Section~\ref{open_challenges}.  

\begin{figure}[!t]
\centering
\includegraphics[width=0.72\linewidth]{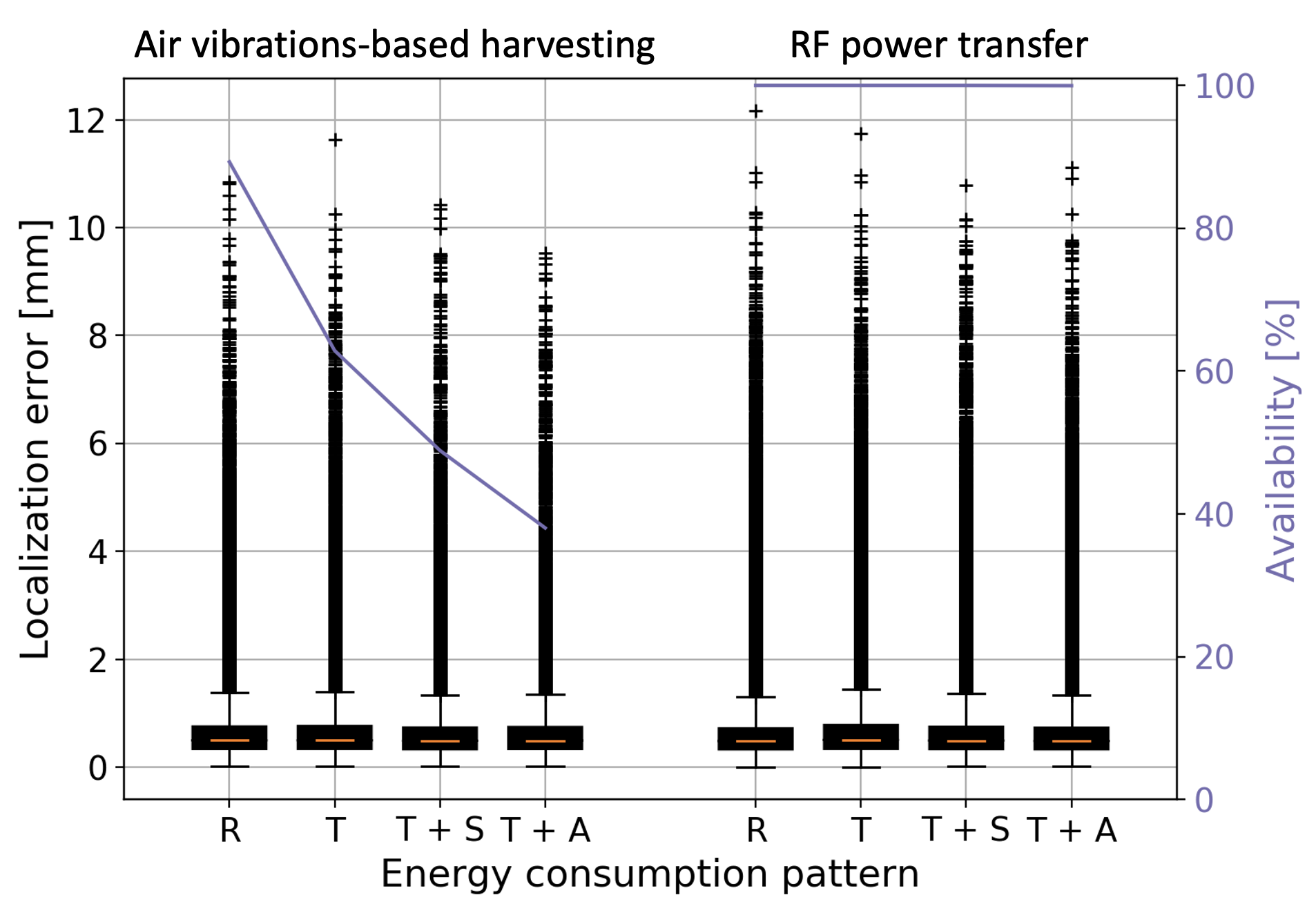}
\caption{Accuracy and availability vs. energy harvesting and consumption profiles (R: receive-only; T: transmit-only; T+S: transmission and sensing; T+A: transmission and actuation)}
\label{fig:consumption_harvesting}
\vspace{-1mm}
\end{figure}

The above results have been derived assuming the metamaterial elements are utilized as instruction-receiving entities only, with the energy in their operational phase being consumed for receiving an 8 bits long packet. 
Additionally, we have assumed that the nanonodes are powered by harvesting the environmental energy from air-vibrations.
In Figure~\ref{fig:consumption_harvesting}, we evaluate the effects of different energy harvesting and energy consumption options on the accuracy and availability of localization in SDMs. 
Specifically, we consider RF-based power transfer as a harvesting option, in addition to air-vibrations. 
This has been done as the SDMs are envisioned to be used for manipulating RF signals, hence harvesting the energy from such signals becomes a feasible and intuitive aim. 
Moreover, we consider the nanonodes being used for transmitting packets to controllers, as well as the combination of sensing/actuation and transmission.
These have been selected as they reflect the SDM usage patterns outlined in the literature~\cite{lemic2019survey,abadal2020programmable}.
The transmission-only pattern has been modeled by assuming that a nanonode transmits an 8 bits long packet in its operational phase.
For the transmission and sensing/actuation-focused patterns, we assumed the energy is consumed in transmission (modeled in the same way as before), as well as sensing/actuation.
The energy consumed in, respectively, sensing and actuation is assumed to be two times lower than and equal to the energy consumed in transmission of an 8-bit packet, which is in line with the existing literature~\cite{canovas2018nature}.

As shown in Figure~\ref{fig:consumption_harvesting}, all usage patterns of a nanonode can be supported in case of RF power transfer. 
This is indicated by the fact that there is no change in the availability of the localization service for this energy harvesting options, suggesting that the nanonodes always have enough energy to perform localization.
However, in case of air-vibrations, which is a significantly weaker harvesting option than RF-based power transfer, the localization service cannot be fully available, regardless of the operational pattern of the nanonodes, which is a direct consequence of a weak harvesting process.   
Moreover, the availability experiences a downward trend as the energy consumed by the nanonodes in their operational phase increases. 
For example, the availability drops from roughly 90 to 40\% if the nanonodes are utilized for actuation and transmission, instead for reception-only purposes. 
The above observations shows that in the design of SDM localization one should account for the energy consumption patterns of the nanonodes, as well as for the amounts of energy the nanonodes are expected to harvest. 
If these considerations result in imperfect availability of the localization service, as it is the case for air-vibrations-based harvesting in Figure~\ref{fig:consumption_harvesting}, one should further consider the requirements from the application(s) to be supported. 
If the application prioritizes the availability over the localization frequency, one should increase the location update period.
This would allow more time for energy harvesting and result in the increased availability of the service, as demonstrated in Figure~\ref{fig:loc_frequency}.  

Also note that, although we have considered that the nanonodes utilize (i.e., transmit or receive) an 8 bits long packet for communication in their operational phases, the results depicted in Figure~\ref{fig:consumption_harvesting} can easily be translated to packets with more or less bits. 
For example, the transmit-only results in the figure are depicted for the operational phase in which the nanonodes transmit 8 bits long packets. 
The same characterization would follow for the reception of 80 bits long packets, given that the nanonodes consume an order of magnitude less energy in the reception than in the transmission of a TS-OOK pulse.  
For this reason, we argue that the performance in case of larger packets in different operational states can be directly interpolated from the results depicted in Figure~\ref{fig:consumption_harvesting}. 
Thus, the localization accuracy and availability as a function of a number of bits in such a packet were not considered.   

Until now, we have assumed that the controllers' locations are known and perfectly accurate. 
As mentioned, we base this assumption on the fact that there are use-cases where an SDM can be mounted on a surface in a way that the controllers' locations are fixed and can be measured.
For the use-cases where that is not the case, we have argued that, given that the controllers are not power-constrained, their locations can potentially be estimated by some other more traditional localization approach or by the same approach as the one utilized for localization of metamaterial elements, only with increased transmit power.  
Whatever the method for localizing the controllers, it will inevitably feature a certain level of localization errors.

\begin{figure}[!t]
\centering
\includegraphics[width=0.72\linewidth]{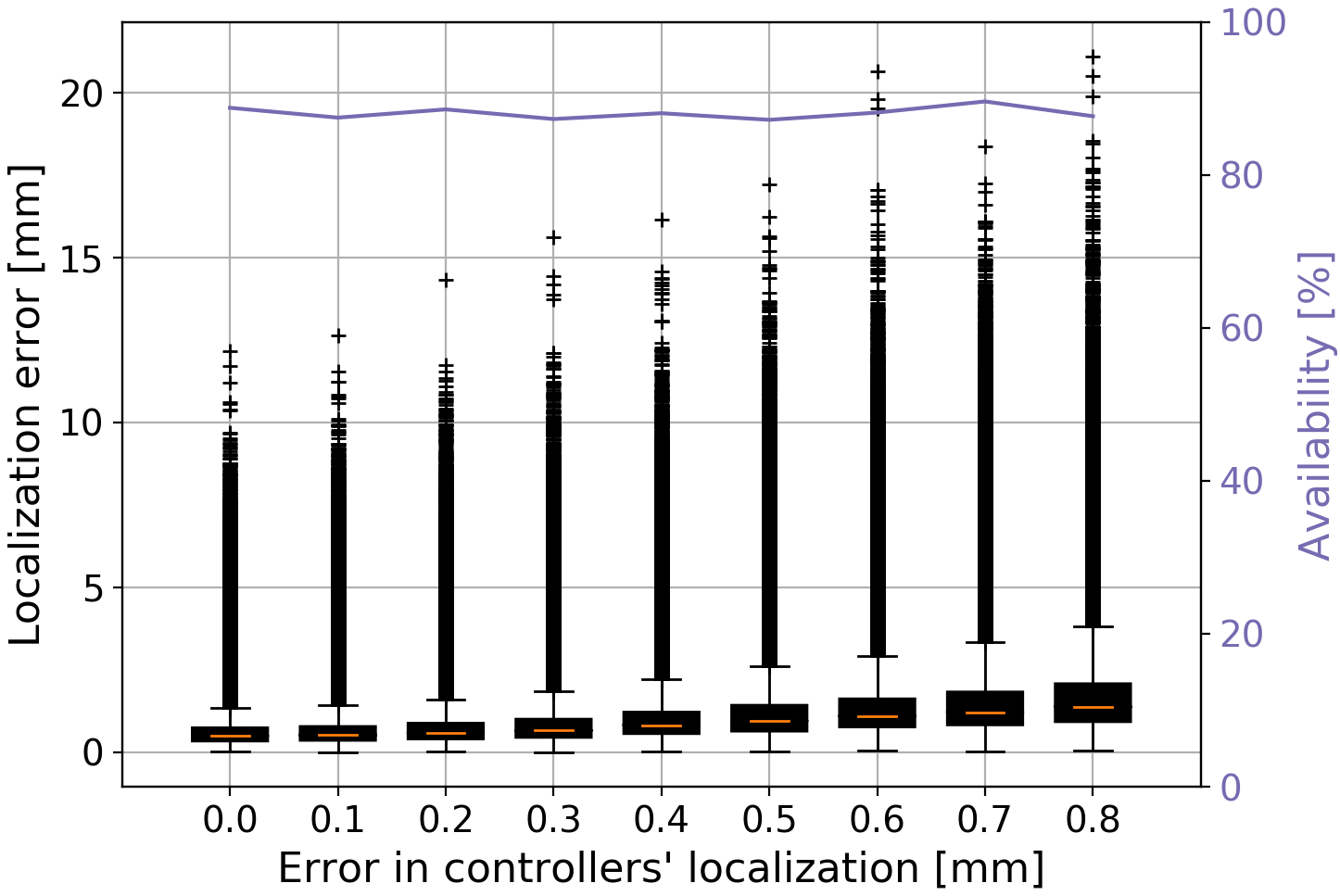}
\caption{Accuracy and availability vs. error in controllers' localization}
\label{fig:error}
\vspace{-1mm}
\end{figure}

In Figure~\ref{fig:error}, we demonstrate the effects of imperfect accuracy of controllers' localization.   
The per-axis localization errors have been modeled as zero-mean Gaussian random variables, which is an established method frequently used in the literature (e.g.,~\cite{behboodi2017hypothesis,lemic2019location}).
Note that in the figure, we depict the \textit{per-axis} standard deviation of the localization error, not the standard deviation of the overall localization error in the controllers' localization.
As depicted in the figure, the accuracy of nanonodes' localization decreases with the increase in the localization errors of the controllers' localization.
Nonetheless, as we have shown before that two-way ToF-based trilateration can achieve sub-millimeter level accuracy of localization of the nanonodes, it is intuitive that comparable (potentially even better) accuracy can be achieved in the controllers' localization.
This is because the controllers are not power-constrained, thus they can utilize higher transmit powers in their localization.
If we assume the average localization error in controllers' localization to be 1.4~mm, which is considerably higher than the one demonstrated for nanonodes' localization, the per-axis error in a 3D space roughly equals 0.8~mm.
As depicted in Figure~\ref{fig:error}, such a per-axis error increases the average localization error from roughly 1 to 2~mm, with most affects being visible in the outliers. 
For example, if the controllers' locations are not perfectly accurate but feature a per-axis error of 0.8~mm, the errors of outliers in SDM localization increase to around 20~mm, in contrast to such errors being consistently lower than 12~mm for perfectly accurate controllers' locations. 
We argue that only a slight increase in the average errors due to the imperfect localization of the controllers shows that the system can operate well even if the controllers are not fixed and their locations are not known with perfect accuracy.

\begin{figure}[!ht]
\centering
\includegraphics[width=0.66\linewidth]{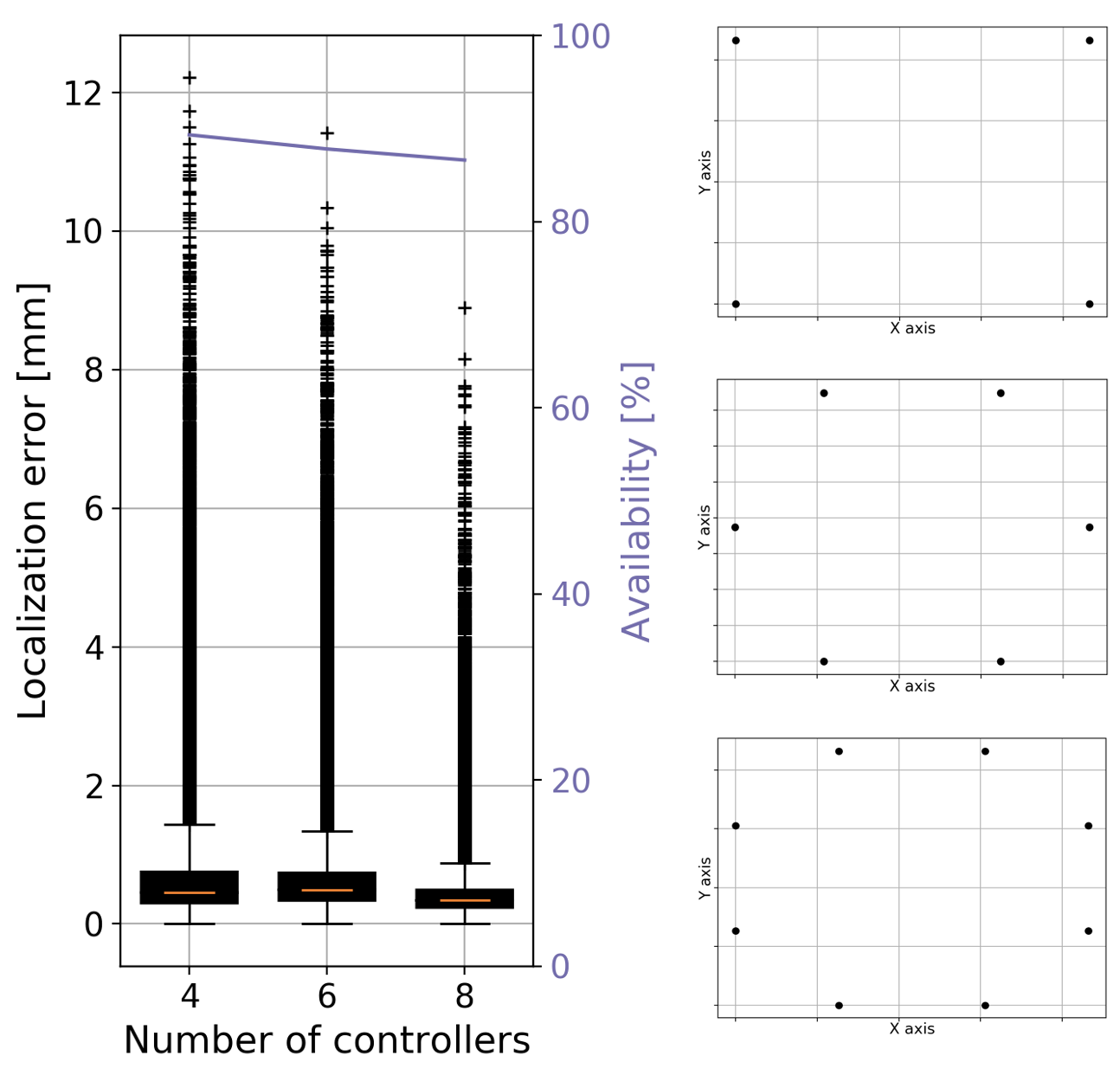}
\caption{Accuracy and availability vs. number of controllers}
\label{fig:anchors}
\vspace{-1mm}
\end{figure}

In Figure~\ref{fig:anchors}, we depict the effects of different number of controllers on the accuracy and availability of localization in SDMs.
As shown in the figure, the accuracy increases with the increase in the number of controllers used as anchors for localization. 
However, this benefit comes at the cost of a more complex design due to the increased number of controllers, as well as an increased latency of SDM localization. 
Also note that, because of the increased chances of nanonodes' power depletion, the availability of localization slightly decreases with the increase in the number of localization anchors.
Similar as with the increasing number of TS-OOK pulses used for two-way ToF estimation (cf., Figure~\ref{fig:nr_pulses}), increasing the number of controllers should be attempted only if the application to be supported requires higher localization accuracy than the one that can be delivered by using the minimum required number of controllers (i.e., four for localization in a 3D space). 

\begin{figure}[!ht]
\centering
\includegraphics[width=0.66\linewidth]{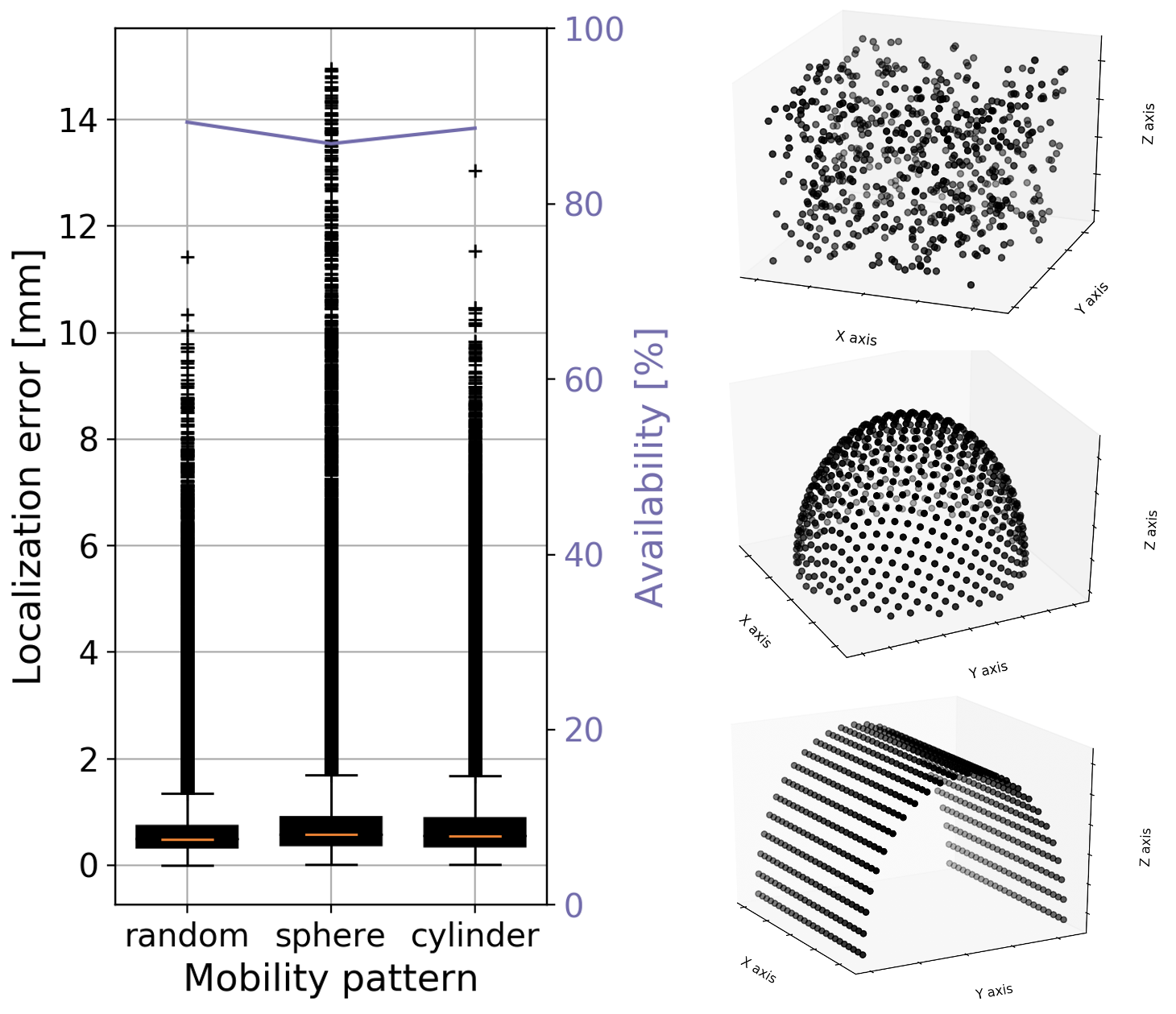}
\caption{Accuracy and availability vs. mobility patterns of the nanonodes}
\label{fig:mobility}
\end{figure}

Finally, Figure~\ref{fig:mobility} depicts the effects of the nanonodes' mobility patterns on the accuracy and availability of localization in SDMs, respectively.
Specifically, the figure shows that the localization accuracy of the outliers decreases in case of cylindrical and spherical mobilities compared to fully random one.  
We argue this can be attributed to the fact that trilateration-based localization in general yields better performance for locations closer to the center of a deployment environment, while the performance degrades closer to the edges of the environment~\cite{lemic2015experimental,van2016performance}. 
In case of the random mobility, more nanonodes are located closer to the center of the localization environment, while for the cylindrical and especially spherical patterns the nanonodes are, generally speaking, located closer to the edges of the environment. 
Note that if the mobility pattern is to an extent predictable, as is the case for the spherical and cylindrical ones, the accuracy of localization could potentially be improved by imposing certain constrains on the nanonodes' estimated locations based on the estimated locations of their neighboring nanonodes. 
Despite the slight increase in the outliers' localization errors for the spherical and cylindrical patterns, the results demonstrate that the proposed approach can work well for different types of nanonodes' mobility.   

\subsection{Latency of Location Estimation}
\label{latency}

From traditional RF-based localization research, it is known that the latency of generating and reporting location estimates is hardware-specific~\cite{dil2010calibration,moayeri2016perfloc,lemic2015experimental,potorti2019evaal}.
Thus, without hardware implementations it is arguably impossible to quantify the latency of localization. 
Nonetheless, qualitative characterizations or trends can still be derived.
Hence, in the following we aim at deriving the qualitative latency of localization as a function of the number of nanonodes to be localized, the number of controllers used for their localization, and the number of TS-OOK pulses used for deriving a single controller/nanonode two-way ToF estimate.

Let us assume that the time needed to obtain a two-way ToF estimate for one controller/nanonode pair equals $t_{ToF}$, while the time needed to perform trilateration for one nanonode once all ToF estimates are obtained equals $t_{tr}$.
Moreover, let $m$, $n$, and $k$ be the number of nanonodes, controllers, and TS-OOK pulses used for two-way ToF estimation for one controller/nanonode pair. 
The latency of localization then consists of the time needed to obtain two-way ToF estimates for all controller/nanonode pairs, as well as the time needed to perform trilateration, multiplied by the number of nanonodes to be localized. 
In the first approximation, we can assume that the time to perform trilateration for one nanonode is constant and does not depend on the number of controllers or TS-OOK pulses used in localization. 
We consider this to be a reasonable assumption, given that this computation time is in practice much lower than the time needed for obtaining the necessary signal features~\cite{wirstrom2015localization,lemic2014experimental}, in this case two-way ToF estimates.  
Under these assumptions, the latency $t_{loc}$ of localization of all nanonodes whose location is to be estimated can be written as:

\begin{equation}
t_{loc} = m (n k t_{ToF} + t_{tr}). 
\end{equation} 

\begin{figure}[!ht]
\centering
\includegraphics[width=\linewidth]{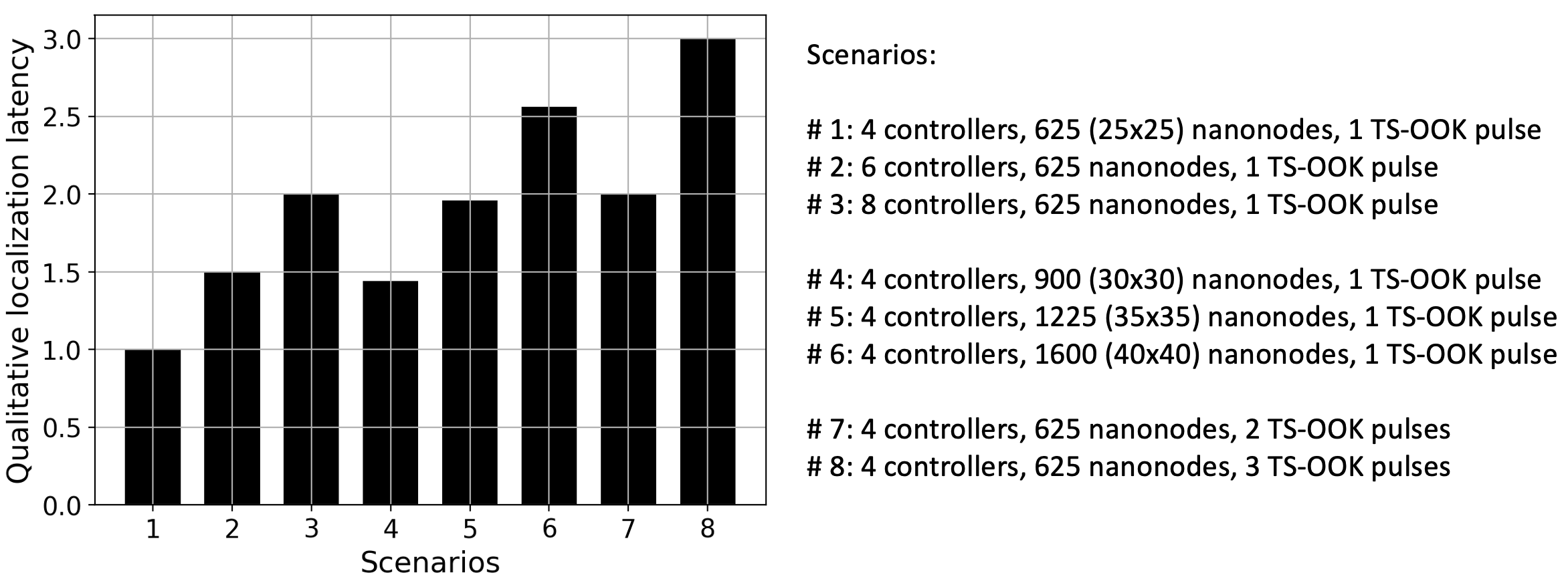}
\caption{Qualitative characterization of localization latency}
\label{fig:latency}
\vspace{-5mm}
\end{figure}

As expected, the localization latency increases with the increase in the number of nanonodes, controllers, and TS-OOK pulses used for deriving a two-way ToF estimate.   
A graphical representation of the qualitative localization latency characterization is given in Figure~\ref{fig:latency}, assuming $t_{ToF} >> t_{tr}$.
Given that the latency increases linearly with the increase in the number of nanonodes, anchors, and TS-OOK pulses, it is intuitive that the increase in the number of TS-OOK pulses used for estimating two-way ToF for one controller/nanonode pair has the most significant negative effect on the latency of localization.
This is because the increase in the number of TS-OOK pulses scales up the latency proportionally to the number of controller/nanonode pairs.
Adversely, increases in the number of controllers and nanonodes affect the localization latency in proportion to the number of nanonodes and controllers, respectively. 
Thus, in the attempt to increase the accuracy of localization, one should first aim at increasing the number of controllers, followed by increasing the number of TS-OOK pulses used for deriving one two-way ToF estimate.
One should also keep in mind that the latency of localization will increase with the increase in the number of nanonodes.	
In order to maintain the localization latency below a certain threshold if an application to be supported requires that, one could aim at increasing the number of SDM subsystems (one such subsystem depicted in Figure~\ref{fig:architecture}) instead of increasing the number of nanonodes in one subsystem. 

\section{Open Challenges}
\label{open_challenges}

Again grounded in the novelty of localization in SDMs, we consider it beneficial to provide our perspective on the remaining open challenges in the realization of such systems. 

\textbf{Novel Use-Cases:} Although we have outlined several high-level example use-cases where localization in SDMs could be beneficial, we believe a more comprehensive and detailed overview is needed.
Providing an exhaustive overview of the potential benefits would make the topic more tangible and, therefore, more attractive to the wider research community. 
By stating ``providing a more detailed overview'' we primarily refer to an overview that specifies the performance requirements that the use-cases pose on the underlying localization service. 
This would address the problem that we will illustrate with the following example.
In this work, we have demonstrated a sub-millimeter achievable accuracy and relatively high availability of the proposed localization service.
However, at this point we cannot provide scientifically accurate statements about the sufficiency of such a service in enabling all the envisaged use-cases, or any for that matter.
Hence, the specifications of the potential use-cases should be detailed in the sense that they quantitatively characterize the requirements that a certain use-case poses on the underlying localization service.       

\textbf{System-level Perspective:}
A system-level solution for localization in SDMs is also required, which could be designed by specifying the signaling requirements and protocols. 
For example, we assumed that two-way ToF measurements can be derived for each controller-nanonode pair.
In reality, deriving such measurements will require a protocol that defines a sequence according to which the nanonodes should transmit signals for ToF estimation.
In this direction, one should realize that, although the SDMs will support structural deformations, the network structure will not change, apart from the nanonodes potentially going out of range or turning off due to the energy depletion.
In other words, the controllers will always be controlling the same sets of nanonodes (i.e., metamaterial elements). 
This suggests that the two-way ToF estimation between a controller and the nanonodes could be done sequentially, potentially even without the need for physical-layer preambles and MAC-layer headers.
We consider the evaluation of this hypothesis to be an open challenge.

We also believe the time needed for obtaining the required ToF measurements could be significantly reduced along two main directions.
First, by properly designing the system one could potentially parallelize the ToF measurements from the standpoint of the controllers, i.e., all controllers could at the same time be performing ToF estimations if there is no interference between the one-controller subsystems. 
Second, one could also embed ToF estimations from a number of nanonodes in a fixed time interval $\beta$ between transmission of two consecutive TS-OOK pulses (or silences).
Specifically, in TS-OOK the time between transmissions is fixed and much longer than the pulse duration (usually two to three orders of magnitude)~\cite{jornet2014femtosecond}, which is a feature originally designed so that the communicating devices do not have to be tightly synchronized, as well as to reduce the effects of interference from collocated transmissions. 
By carefully designing the reactions of the nanonodes to the pulse transmitted by the controller, one could achieve a number of ToF estimations by transmitting only a single TS-OOK pulse on the controller level. 
Note that such an approach would also benefit the energy efficiency of the solutions, as the nanonodes would consume energy for receiving only one TS-OOK pulse, in contrast to a number of them if the estimation is done by sending a pulse by the controller for each controller/nanonode pair. 

\begin{figure}[!t]
\centering
\includegraphics[width=0.6\linewidth]{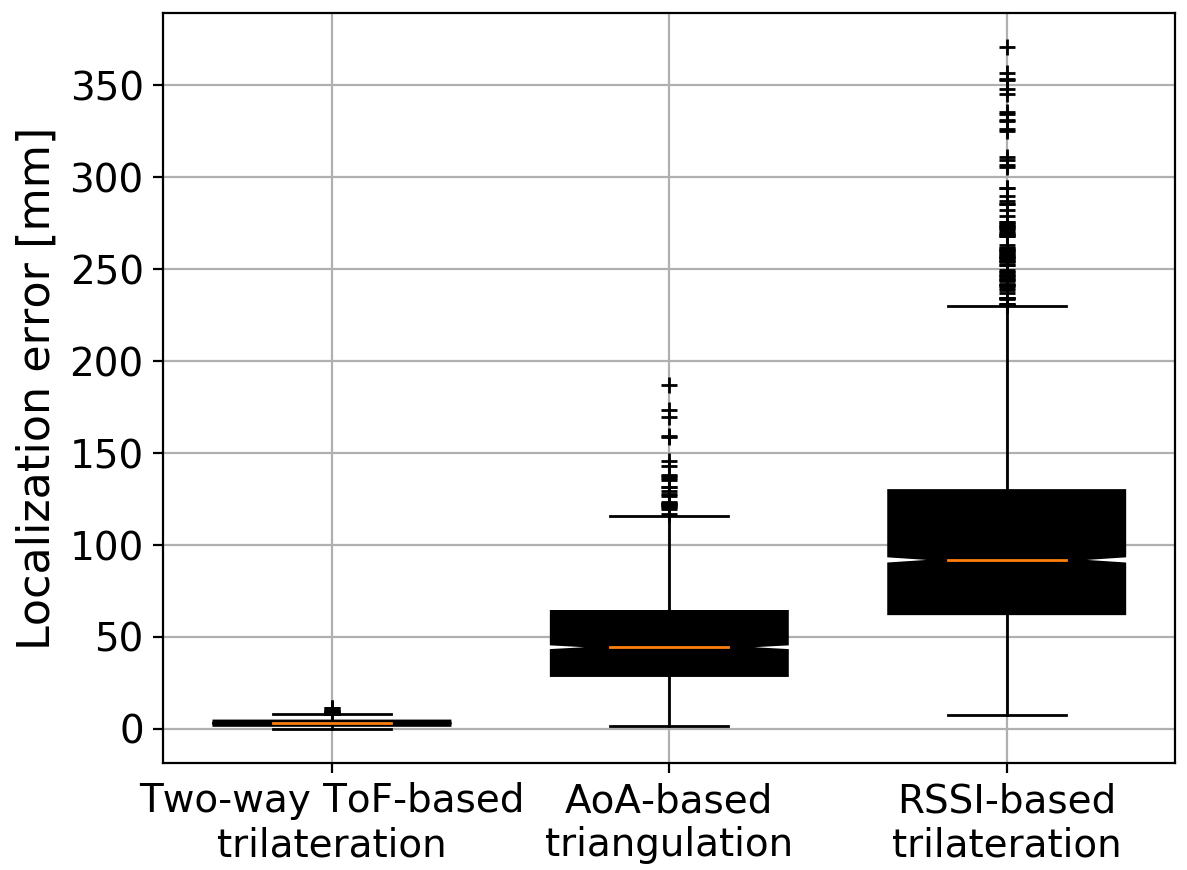}
\caption{Localization accuracy in comparison to AoA-based triangulation and RSS trilateration-based localization}
\label{fig:comparison}
\end{figure}

\textbf{Localization Approaches:} In this work, we have considered the main instances of traditional RF-based localization approaches and argued that two-way ToF-based trilateration is the most suitable among them for enabling localization in SDMs.
In other words, we believe that the traditionally utilized alternatives, primarily \ac{AoA}-based triangulation and \ac{RSS}-based trilateration approaches, cannot be utilized for localization in SDM in the near future due to the specifics of the nanoscale scenario at hand.
In a hypothetical example of achieving the AoA resolution of 0.75 degrees and RSS variability characterized with a zero-mean Gaussian distribution with standard deviation of 0.3~dB, one would be able to localize the SDM elements with the accuracies as shown in Figure~\ref{fig:comparison}.
Note that the selected resolution and variability values are significantly lower than what the state-of-the-art macro-scale THz systems are able to achieve~\cite{aghoutane2021spatial,peng2019power}.
Comparable AoA resolutions and RSS variabilities at nanoscale are, therefore, hardly achievable in the near future, hence we portray these results only as a hypothetical example. 
In addition, in the derivation of the results depicted in the figure, we did not consider the energy consumption of each of the approaches, but assumed that the nanonodes have sufficient energy for localization.
This is mainly due to inability of modeling energy consumption in AoA-based triangulation, given that such a system intrinsically cannot be built assuming omnidirectional TS-OOK-based nanoscale communication.
The other simulation parameters are consistent for all three approaches from Figure~\ref{fig:comparison} and summarized in Table~\ref{tab:paramaters}.
As depicted in Figure~\ref{fig:comparison}, the ToF-based localization yields more than an order of magnitude lower localization errors than the two considered alternatives.
This is a direct result of a large available bandwidth at THz frequencies, which, we believe, makes an argument for further developments of ToF-based localization approaches for THz-operating TS-OOK-based nanoscale communication systems.

However, in our evaluation we have assumed large channel bandwidth (implying extreme sampling rates and fine-grained time discrimination on the level of a nanonode) that are hard to achieve in practice. 
We have done this because our aim was to demonstrate the potential of localization in SDMs by showing its theoretical performance limits.
Nonetheless, there are several approaches in the existing literature aiming at generating accurate two-way ToF measurements under limited sampling frequencies~\cite{lanzisera2011radio,wirstrom2015localization}, as well as for tight pulse-based synchronization at the nanoscale~\cite{gupta2015joint,xia2019link}.
The applicability of such approaches for the problem at hand is yet to be established.   

Given that the traditional RF-based localization research is at this point abundant, we believe there are other lessons to be learned from those efforts. 
Hence, more advanced SDM localization approaches could be developed, which would in turn further advance the performance of localization in SDMs. 
One approach we have considered is multi-dimensional scaling~\cite{saeed2019state}, a localization approach that assumes the existence of an interconnected graph between (nano)nodes.
Based on the graph, the approach is able to estimate the locations with higher accuracy than regular trilateration~\cite{wei2009multidimensional}, and it works particularly well for structured grid-like graphs, hence it is intuitively suitable for SDMs.   
However, we have observed that the energy consumed in the process of obtaining and delivering all ToFs (distances) between connected nanonodes to the controller level makes the approach infeasible for the primarily considered nanonodes with air vibration-based energy harvesting. 
Nonetheless, this does not imply that other approaches could not advance the currently reported performance metrics.

One could also consider entirely replacing RF-based localization with other types of approaches.
For example, a inductive coupling-based localization approach has been shown to yield promising performance at the microscale~\cite{slottke2016inductively}. 
However, it is currently unclear if such a solution can be further downscaled and if the power consumption of such a miniaturized solution would be acceptable from the perspective of energy harvesting nanonodes.

\textbf{Anchor/controller Localization:} In this work, we have assumed the localization anchors are either fixed or their locations can be estimated with high levels of accuracy. 
Although we believe the assumption of the anchors' locations being fixed, therefore measurable and perfectly accurate, is feasible for some use-cases such as e.g., structural monitoring, for others there will certainly be a need for estimating the locations of the anchors. 
The question on how to perform such localization is currently not clear, given that in order to localize the nanonodes, the accuracy of the anchors' localization should be higher than what the existing approaches (mostly using sub-6~GHz frequencies) can achieve.
We believe the answer here lies in the utilization of high frequency and high bandwidth localization approaches (i.e., mmWave and THz) that show a high promise in enhancing the accuracy of localization~\cite{lemic2016localization,palacios2019single}.
However, the question of the feasibility of such approaches is, to the best of our knowledge, still open, especially for the challenging environments (e.g., smart textiles worn by a person) relevant in this context.   

\textbf{Prototypical Implementations:} By modeling and simulating a new localization solution one can obtain initial insights on its performance in realistic deployments. 
For example,~\cite{lemic2016localization} is an early work on mmWave macro-scale localization, where the authors utilize a simulation-based approach for demonstrating that trilateration-based localization in mmWave frequencies can achieve the accuracy on a decimeters-level.
The utilized simulation approach is similar in its degree of realism to the one utilized in this work.
These initial implications have been experimentally confirmed by~\cite{palacios2019single}, where the authors demonstrated the average localization error of around 30~cm for ToF and AoA-based trilateration.
Prototypical implementations of localization-enabled THz-operating power-constrained SDMs will be needed for confirming (or discarding) the insights made in this and potentially future simulation-oriented works.  
In this direction, pioneering efforts on developing graphene-based antennas~\cite{jornet2013graphene, singh2020design} and transceivers~\cite{akyildiz2016graphene,cid2012receiver}, micro and nanoscale energy-harvesters~\cite{murillo2011hybrid,lozano2020non,rahmani2018dual,klinefelter201521}, as well as SDMs prototypes~\cite{liaskos2018new,zhang2018space,yang2016programmable,tran2019novel,dai2020reconfigurable,kossifos2020toward,liu2019intelligent} are worth emphasizing.

\textbf{Performance Evaluations:} Similar as in traditional localization utilizing RF signals, in the context of nanoscale localization in SDMs multiple performance metrics will be needed to fully characterize the performance of newly proposed localization systems. 
For example, the EVARILOS Benchmarking Handbook~\cite{van2013evarilos} provides the specifications of an exhaustive set of such metrics, encompassing among others the localization accuracy, reliability (i.e., availability), latency, energy efficiency, deployment overhead, scalability, etc.
In addition, the handbook aims at standardizing the experimentation conditions required for the benchmarking of new solutions, so that the obtained benchmarks can later on be compared among each other in a meaningful way. 
In the context of nanoscale localization in general, such specifications and benchmarking guidelines do not exist, which could result in incomparable performance results with low utility. 

In the consequent step, experimental benchmarking should ideally be performed under the same conditions (e.g., same environment, same hardware, evaluation points), which is an issue that has been recognized in the community focused on traditional RF-based localization, resulting in a number of localization competitions and similar events aiming at comparative benchmarking of localization solutions (e.g.,~\cite{lymberopoulos2015realistic,lemic2015experimental,potorti2019evaal}).
Consequently, the data obtained in the experimentation started becoming available and even accessible through web services and platforms, so that the only thing an experimenter has to do is to upload the code of a localization solution, while the service deals with feeding it unified experimental data, gathering and processing of the outputs provided by the solution, and positioning of the solution in the context of already benchmarked ones~\cite{van2015platform,moayeri2016perfloc,moayeri2018perfloc,lemic2015web}.
Similar tools and methodologies are needed for nanoscale localization to avoid the ``status quo'' where various parties claim the general superiority of their localization solution based on incomplete and incomparable performance benchmarks. 

Finally, if the applications require that the localization errors of all estimates are below a certain threshold, one could attempt to develop a system for estimating localization errors on a per-estimate basis, followed by discarding the estimates whose errors are predicted to be higher than the threshold.
In this direction, it is worth emphasizing~\cite{lemic2019regression}, where the authors utilize machine learning for solving this problem. They show that the system can achieve the accuracy of estimation of localization errors of an per-estimate basis that is significantly higher than the one that can be obtained through static benchmarks. Nevertheless, the proposed system is focused on the macroscale, considers a different localization approach (i.e., fingerprinting vs. trilateration) and different networking technologies (i.e., SigFox and LoRa vs. in-body nanonetwork), among others. Hence, the utility of such or similar systems for the considered setup is yet to be established.

\section{Conclusion}
\label{conclusion}

We have shown that the two-way \acf{ToF}-based trilateration has a potential for enabling accurate localization in \acf{THz}-operating and power-constrained \acfp{SDM}. 
We base our indication on the sub-millimeter accuracy and high availability of localization for the system parameterizations expected in real-life SDM implementations. 
Moreover, we have qualitatively characterized the effects of several relevant system parameters.
Example-wise, we have shown that the utilized bandwidth significantly affects the localization accuracy, while the energy harvesting rate and location update period play an important role in the service availability. 
Finally, we have shown that the proposed localization solution can be introduced to energy-harvesting (and therefore power-wise highly constrained) nanonodes without incurring significant energy-related constraints to the operation of the nanonodes. 

In conclusion, we argue that this work provides a new argument for the utilization of THz-based wireless solutions for the control and programming of SDMs, in contrast to the utilization of wired solutions.
This is because a wireless solution, in addition to being form-factor-wise superior, would allow localizing the metamaterial elements under mobility, which in turn could be an enabler of novel applications, as well as novel approaches for location-based optimization of the nanonetwork for controlling SDM elements.  
Future work will focus on addressing some of the challenges outlined in this paper.

\section*{Acknowledgments}
This work was supported by the EU Marie Curie Actions Individual Fellowship (MSCA IF) project Scalable Localization-enabled In-body Terahertz Nanonetwork (SCaLeITN), grant nr. 893760. 
Filip Lemic’s research stay at the Terahertz Wireless Communications Lab at Shanghai Jiao Tong University, China was supported by the Research Foundation - Flanders (FWO) in the form of a grant for a long stay abroad (grant nr. V406320N). 
This work also received support from the European Union via the Horizon 2020 Future Emerging Topics call (FETOPEN, grant no. 736876).

\section*{Bibliography}
\label{}


\bibliographystyle{elsarticle-num}

\end{document}